\documentclass[namedreferences]{solarphysics}

\usepackage[hyperref,optionalrh]{spr-sola-addons} 
\usepackage{graphicx}        
\usepackage{amssymb}         
\usepackage{amsmath}         
\usepackage{color}           
\usepackage{breakurl}       
\usepackage{hyperref}
\hypersetup{
    colorlinks=true,
    linkcolor=blue,
    filecolor=blue,      
    urlcolor=blue,
}

\usepackage{verbatim}
\usepackage{float}
\usepackage{enumitem}
\usepackage{mathtools}

\chardef\us=`\_

\newcommand{\clark}{CPP}

\begin{document}
\begin{article}
\begin{opening}

\title{Automated Spatiotemporal Analysis of Fibrils and Coronal Rain using the Rolling Hough Transform}

%
\author[corref,addressref=aff1,email ={schad@nso.edu}]{\inits{T.A.}\fnm{Tom}~\lnm{Schad}}  
\address[id=aff1]{National Solar Observatory, 8 Kiopa`a Street, Pukalani, HI 96768, USA}
\runningauthor{T.A. Schad}
\runningtitle{Rolling Hough Transform Analysis of Solar Structures}


\begin{abstract}
A technique is presented that automates the direction characterization of curvilinear features in multidimensional solar imaging data sets. It is an extension of the Rolling Hough Transform (RHT) technique presented by \cite{clark2014}, and it excels at rapid quantification of spatial and spatiotemporal feature orientation even for low signal to noise applications. It operates on a pixel-by-pixel basis within a data set and reliably quantifies orientation even for locations not centered on a feature ridge, which is used here to derive a quasi-continuous map of the chromospheric fine structure projection angle.  For time-series analysis, a procedure is developed that uses a hierarchical application of the RHT to automatically derive apparent motion of coronal rain observed off-limb. Essential to the success of this technique is this paper's formulation for the RHT error analysis as it provides a means to properly filter results.
\end{abstract}

\keywords{Active Regions: Structure $\cdot$ Corona: Structures $\cdot$ Chromosphere: Active $\cdot$ Methods: Pattern Recognition}

\end{opening}


\section{Introduction} \label{sec:intro}

On account of its magnetic field, the solar atmosphere contains curvilinear structures often spoken about in terms of loops and their likely, perhaps complicated, association with lines of magnetic force \citep[see, \textit{e.g.},][]{delacruz2011, schad2013, leenaarts2015, chen2015}.  Based on that association, the presence of coronal loops, fibrils, and spicules help derive information about the solar magnetic field, and thus methods are required to extract their geometrical attributes. 

In many cases, it is only necessary to extract the projected orientation of magnetized features.  For example, the projected orientation of coronal and chromspheric features help constrain non-linear force free modeling of coronal fields, as discussed by \cite{wiegelmann2008} and \cite{aschwanden2016}.  Measuring the misalignment angle between chromospheric fibrils and potential field extrapolations may provide a quick means to judge magnetic free energy in active regions \citep{jing2011}.  Furthermore, knowledge of feature orientation may aid in the inversion of chromospheric spectropolarimetric data used to infer the chromospheric magnetic field, in particular to manage the role of ambiguities and allow inference of other radiatively controlled variables such as material height \citep[see Section 6.2 of][]{asensio_ramos2008}.  

The reliable determination of the projected orientation (or direction) of solar curvilinear features in regimes with many hundreds or thousands of features require deterministic mathematical-based processing techniques.  Many pattern recognition techniques have been developed in solar physics and were comprehensively reviewed by \cite{aschwanden2010}.  For coronal loops, most techniques concentrate on tracing the coordinates of 1-D image \textit{ridges}, \textit{i.e.} a set of local \textit{maxima} forming a one-dimensional locus, from which the projected orientation derives. This approach necessarily assumes each analyzed features has a single well defined ridge.  Prominent methods include the oriented-connectivity method \citep{lee2006_oriented}, the dynamic aperture-based loop-segmentation method \citep{lee2006_aperture}, the unbiased detection method of \cite{steger1996}, the oriented directivity method (or OCCULT, which stands for the \textit{Oriented Coronal CUrved Loop Tracing}) of \cite{aschwanden2010_occult} \citep[see also][]{aschwanden2013} , and ridge detection by automated scaling \citep{inhester2008}, all of which were compared by \cite{aschwanden2008_compare}.  Of these, OCCULT typically returns the largest number of features in coronal applications and rivals manual-based analysis.  

Local orientation/direction analysis is one key step in OCCULT. As described in \cite{aschwanden2010_occult}, OCCULT connects points along individual ridges by bidirectional step-wise tracing along the direction aligned with the ridge axis.  The axis is determined by integrating the measured flux in a ridge-enhanced (\textit{i.e.}, high and lowpass filtered) image along curve segments of a given direction angle and radius of curvature anchored on a single point and discretized for all angles and user-specified curvature radii.  The segment with the maximum integrated flux is selected as the ridge axis.  While quite successful at tracing coronal loop ridges, this method is sensitive to a few assumptions.  It relies on the anchor pixel being located on the ridge itself.  Similarly, it depends on the ridge being analyzed not lying close to a feature with significantly greater flux.  Should these cases arise, the integrated directed flux measured identifies a direction not aligned with the ridge but instead in the direction between the point and the region of greater flux.  The method further assumes that the integrated flux function has at least one well defined peak, meaning that the curvilinear features needs to have a well defined cross-section with a local maxima.  Thick features observed with low signal-to-noise do not always give a well defined peak.  Finally, the method does not infer the local direction for pixels not lying on a ridge.  

An accurate method for local orientation/direction analysis that overcomes these limitations has specific applications.  In the case of the chromosphere, ridge-like fibrils that give the chromosphere its dense fine-structure contrast are likely signatures of thermal and density perturbations within a magnetic field with a spatially varying magnitude that is much smoother by nature.  Some evidence of this is shown by \cite{schad2015}.  Thus, there is an advantage to developing a technique that determines the orientation of thick and densely populated curvilinear features that need not exhibit a well-defined ridge.  Moreover, such a technique should determine the orientation for pixels between the ridges and/or the central feature axis reliably without resorting to extrapolation of information derived along a 1-D ridge/locus.  Such information can help inform both field modeling and spectropolarimetric inversions.

A new technique could also potentially automate traditional time-slice analysis of apparent motion along curvilinear trajectories.  For a number of applications, motion along loops instantiate 1-D curvilinear features traversing both space and time that might be analyzed with a local direction analysis in three-dimensions to extract the apparent velocity (speed and direction).  A key motivator here is establishing a method to quantify the velocity of a greater number of cool blobs that form and exhibit motion during coronal rain events.  The most extensive analysis of fine-scaled coronal rain velocities to date resulted from the tedious manual time-slice analysis conducted by \cite{antolin2012sharp}.  Coronal rain is a dim and rapidly evolving phenomena typically measured with low signal-to-noise, i.e. a few counts above the background.  Cool blobs can exist in close proximity to each other while exhibiting significantly different velocities and brightnesses that evolve with time independently.  Such characteristics are a challenge for existing automated feature tracking and/or optical flow algorithms, such as local correlation tracking \citep{november1988,welsch2004} and Lucas-Kanade spatial derivative techniques \citep{lucas1981,gissot2007}.
 
Here a robust technique for local orientation analysis is developed by extending the Rolling Hough Transform (RHT) introduced by \cite{clark2014} (hereafter \clark), which has already been applied to images of chromospheric fibrils by \cite{asensio_ramos2017_fibril}.   Section~\ref{sec:2DRHT} describes the RHT procedure and develops an approach to error analysis based on circular statistics.  Numerous illustrations are given to demonstrate the technique's performance.  Section~\ref{sec:rht_howto} outlines how to apply the RHT procedure to solar data.  Section~\ref{sec:rht_trace} applies the technique on the same sub-image of coronal loops used by \cite{aschwanden2008} to compare coronal loop tracing codes.  Section~\ref{sec:rht_ibis} extends the technique to derive a quasi-continuous map of the azimuthal orientation of fine structure in the chromosphere.  Finally, Sections~\ref{sec:time_slice_rht} and~\ref{sec:rht_iris} extend the technique to the time-slice motion problem with application to coronal rain observations.  


\section{Two-Dimensional Rolling Hough Transform Analysis} 


\subsection{RHT definition}  \label{sec:2DRHT}

The Rolling Hough Transform, introduced by \clark, derives its name from the Hough Transform developed for the automated ``machine" analysis of subatomic particle tracks in bubble chamber photography \citep{hough1959, hough1962}.  Hough's technique divided images consisting of numerous discontinuous curvilinear bubble tracks into``framelets" wherein the curve segments could be approximated as lines. Instead of searching for sets of points with linear correlation, Hough transforms each bubble track point within a given framelet into a 2D parameter space representing the slope and intercept of all possible lines in free space.  The transform maps each point to the set of all lines that pass through it, and thereby acts as a voting process where each point in the image adds a vote to a set of candidate lines in the parameter space, which is often called the ``accumulator."   Peaks in the accumulator identify the parameters of the lines in the image, which need not be continuous.  This powerful technique can be generalized to other geometries \citep{mukhopadhyay2015}.  For lines it is numerically preferred to use the normal form parameterized by its normal angle ($\theta$) and Euclidean distance from the origin ($\rho$), as in \cite{duda1972}. 

\clark\ modified the Hough transform technique by creating a localized `rolling' version that acts on individual points within an image that is either binary (values of 0s and 1s) by nature or has been binarized via some segmentation method.  The procedure is equivalent to performing a Hough transform within a circular domain of a given size, centered on an individual pixel, and restricted to lines with a zero Euclidean distance ($\rho$), \textit{i.e.}, those lines passing through the center pixel.  Stated alternatively, within a circular kernel centered on point $(x,y)$, it adds up the number of "illuminated" (\textit{i.e.}, value of 1) pixels along all axes\footnote{The term \textit{`axis'} signifies that the parallel and anti-parallel directions are treated equivalently, meaning a given direction is sampled along a line extending through the origin and between two points on the kernel's edge.  Thus, the data measures non-vectorial undirected lines.}  surrounding the center pixel.  For the ideal case of a thin straight line, the returned function $H_{xy}(\theta)$ is peaked at the angle corresponding to the local direction associated with the linear feature traversing the center pixel.  In this way, it is very similar to the OCCULT directivity method by  \cite{aschwanden2010_occult} with the exception that it works on \textit{any} pixel within a \textit{binary} image; it does not require the presence of a ridge\footnote{See definition of ridge in Section~\ref{sec:intro}.} and does not integrate measured flux.  In the examples below,  the advantages of this subtle difference will be demonstrated.

\begin{figure}
\centering
\includegraphics[width=0.95\textwidth]{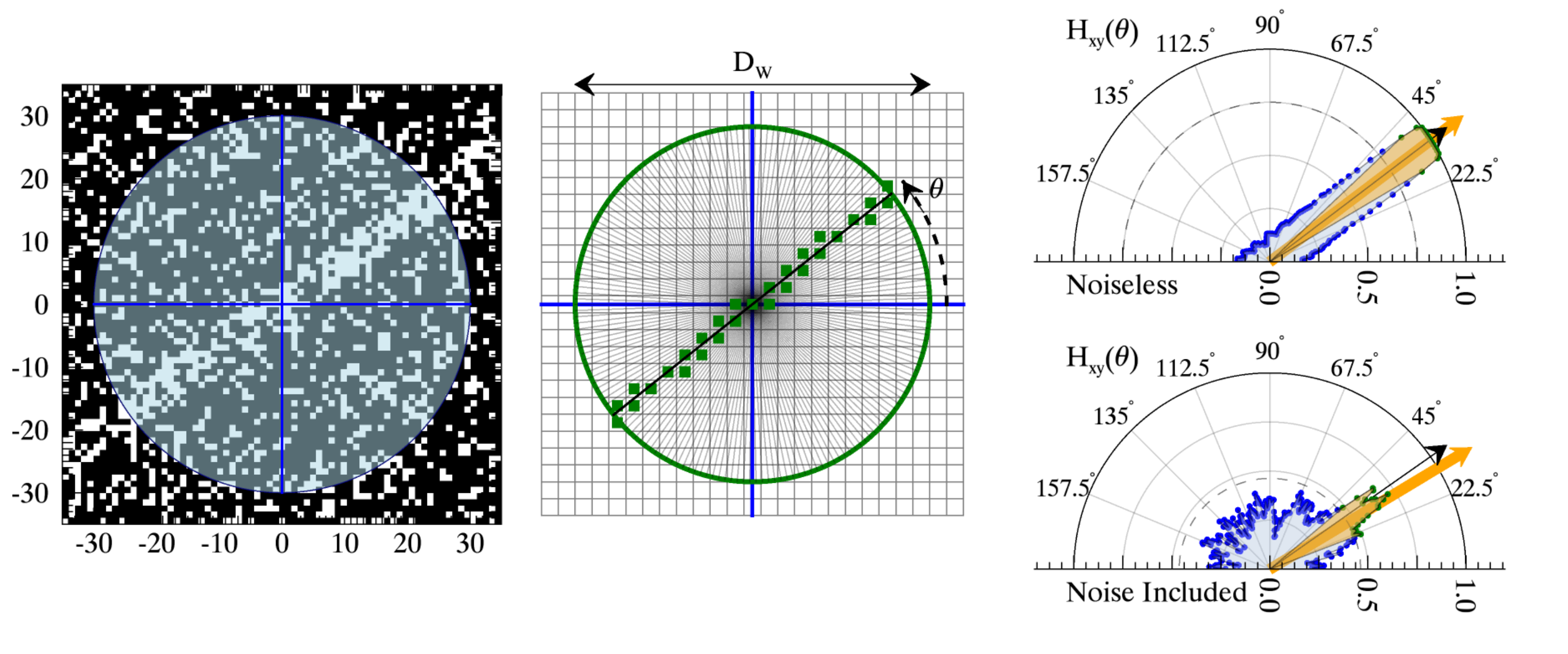}   
\caption{Demonstration of the Rolling Hough Transform (RHT).  (\textit{left}) Synthetic binarized image of a straight line with non-zero thickness and an additive noise contribution.  The blue crosshair intersection denotes the pixel being analyzed by the RHT lying inside of the circular kernel ($D_{W} = 51$) which is indicated by the transparent overlay.  (\textit{middle}) A graphical representation of the RHT identifying all pixels intersecting the axis orientated at an angle $\theta$.  A sum of the binarized image is calculated for all pixels along each $\theta$ and then normalized by the number of pixels.  (\textit{top right}) The RHT transformed function $H_{xy}(\theta)$ for the use case without added noise. (\textit{bottom right}) $H_{xy}(\theta)$ for the use case including noise (as shown in left panel).  The black arrow denotes the known orientation of the line while the orange arrow corresponds to the RHT derived orientation angle.}
\label{fig:simple_demo}
\end{figure}

Figure~\ref{fig:simple_demo} illustrates the RHT for a binary image of a line with a small relative thickness by deriving the function $H_{xy}(\theta)$ for the noiseless case and with noise added.  The circular kernel size has a width of $D_{W}$ pixels (\textit{i.e.,} a diameter of $D_{W} -1$ pixels).  Along each \textit{axis} $\theta$ (sampled as defined below), the number of illuminated pixels that intersect the axis are counted and normalized by the total number of pixels intersecting the axis direction.  In the polar plots in Figure~\ref{fig:simple_demo}, $H_{xy}(\theta)$ ranges from 0 to 1; a value of 1 signifies all pixels in that direction are illuminated.  In both the noiseless and noisy case, the function is peaked along the direction of the line. 

Given this technique, a procedure is needed that accurately defines the mean direction of the rolling Hough transformed function $H_{xy}(\theta)$ and its error.  \clark's approach calculated the circular statistical mean of all values above a user-specified threshold.  While this approach is better than simply taking the maximum, it unfortunately does not adequately quantify the significance of the mean direction.  Cases may exist where only a single orientation $\theta$ has a value above a given threshold but that does not represent a significant peak.  Similarly, if all pixels are illuminated, a mean direction can be calculated but has no significance. 

In the presence of random noise, the circular statistical mean of $H_{xy}(\theta)$ will be negatively influenced by non-isotropic groupings of noise.  Due to this effect, an adaptive threshold is defined here that scales according to the peak of $H_{xy}(\theta)$ such that all values below the threshold do not contribute to the mean direction.  The advantage of this is that values within some range of the peak are still included as a measure of the peak's significance.   This adaptive thresholding can be written as
\begin{equation}
h_{xy}(\theta) =  
\begin{dcases}
	H_{xy}(\theta), & \text{if } H_{xy}(\theta) \geq \max [ H_{xy}(\theta) ] - f \\ 
	0 , &  \text{if } H_{xy}(\theta) < \max [ H_{xy}(\theta) ] - f. \\ 
\end{dcases}
\end{equation}
Here a value of 0.25 is used for $f$.  In other words, a peak needs to be 25\% greater than the surrounding noise to be well isolated by the adaptive threshold.

To adequately define the mean direction and its error, the RHT must critically sample the direction axes within the circular window of width $D_{W}$ for $\theta \geq 0$ and  $\theta < \pi$.  In contrast with \clark, here critically sampling is defined as Nyquist-Shannon sampling of the linear distance around the kernel's circumference, \textit{i.e.} a discrete angle step size (in radians) of 
\begin{equation}
\delta\theta = \frac{1}{2}\frac{2\pi}{\pi (D_{W}-1)}  = \frac{1}{(D_{W}-1)}.   \label{eqn:theta_sampling}
\end{equation}
This approach leads to finer angular sampling than \clark\ but ensures that the mean direction is not overly discretized.  Tests using the canonical sampling of \clark\ resulted in systematic groupings in the calculated mean directions of points.  

Because this is axial data, \textit{i.e.} orientations of $2^{\circ}$ and $178^{\circ}$ are not widely separated, the data is first transformed into vectorial data so that vectorial circular statistics can be used to derive the mean direction.  This method, which uses the approach of doubling all angles, was introduced by \cite{krumbein1939} and was further discussed by \cite{fisher1995statistical}.  \citeauthor{fisher1995statistical} recommends that the mean direction results of such transformed axial data be back-transformed, meaning here that the result is interpreted as an axis rather than a vector. Measures of spread, dispersion, and confidence intervals are left in units of the transformed (vectorial) data. 

Using the doubled angle \textit{axial} form of the direction statistics given by \cite{mardia2009}, the mean axial orientation is calculated by first calculating the weighted Cartesian coordinates of its vectorial counterpart, \textit{i.e.}, 
\begin{align}
\bar{C} &= \frac{ \sum_{\theta} h_{xy}(\theta) [ \cos 2\theta]} {\sum_{\theta} h_{xy}(\theta)} ,  \\
\bar{S} &= \frac{ \sum_{\theta} h_{xy}(\theta) [ \sin 2\theta]} {\sum_{\theta} h_{xy}(\theta)}. 
\end{align}
The mean axial direction\footnote{The factor of 0.5 in introduced by the back-transformation into axial coordinates.} is then given by
\begin{equation}
\bar{\theta} = 
\begin{dcases}
0.5 \cdot \arctan (\bar{S} / \bar{C}) , &  \textit{if } \bar{C} \ge 0,  \\ 
0.5 \cdot \arctan (\bar{S} / \bar{C})  + \pi,&  \textit{if } \bar{C} < 0.
\end{dcases}
\end{equation}
As a measure of concentration, it is useful to define the quantity referred to as the mean resultant length \citep{mardia2009}, which is given by
\begin{equation}
\bar{R} = \sqrt{\bar{C}^{2} + \bar{S}^{2}}
\end{equation}
and corresponds to the length of the average vector pointing along the mean axial direction.  Its value ranges from 0 to 1, where 0 means that $h_{xy}$ is uniformly distributed and the mean axial direction is ill-defined.  $\bar{R}$ is valued at 1 when $h_{xy}$ represents a circular delta function and the mean axial direction is well-defined.  Thus, $\bar{R}$ is a measure of dispersion.  It is related to the circular standard deviation $\sigma$ as \citep{mardia2009}: 
\begin{equation}
\sigma = \sqrt{-2 \ln \bar{R} }.
\end{equation}
An approximate $(1-\alpha)$ confidence interval\footnote{$\alpha$ is the canonically defined significance/confidence level.}, $\bar\theta \pm \epsilon_{\theta}$, for the mean axial direction $\bar\theta$ is defined by \cite{fischer1983}, making no assumptions regarding the underlying shape of the distribution but in the limit of a large number of samples $n$, as follows
\begin{equation}
\epsilon_{\theta} = \sin^{-1} \left (  u_{\alpha} \sqrt{\frac{(1-\alpha_{2})}{2n\bar{R}^{2}}} \right ) \label{eqn:conf_int}
\end{equation}
where $u_{\alpha}$ is the upper $\frac{1}{2}\alpha$ quartile of the N(0,1) distribution and $\alpha_{2}$ is related to the variance of the underlying distribution and is estimated here as 
\begin{equation}
\hat{\alpha}_{2} = \frac {\sum_{\theta} h_xy(\theta) [ \cos 2(\theta - \bar\theta) ] }{ \sum_{\theta} h_xy(\theta)}.
\end{equation}
Using this formulism for the confidence interval negates the need to model the measured distribution thereby greatly minimizing the computational time required to estimate errors. Here the 95\% confidence interval ($\alpha = 0.05 , u_{\alpha} = 1.96$) is reported as the error in the mean axial direction, and $n$ in Equation~\ref{eqn:conf_int} is the number of non-zero values in $h_{xy}$. 

\begin{figure}
\centering
\includegraphics[width=0.31\textwidth]{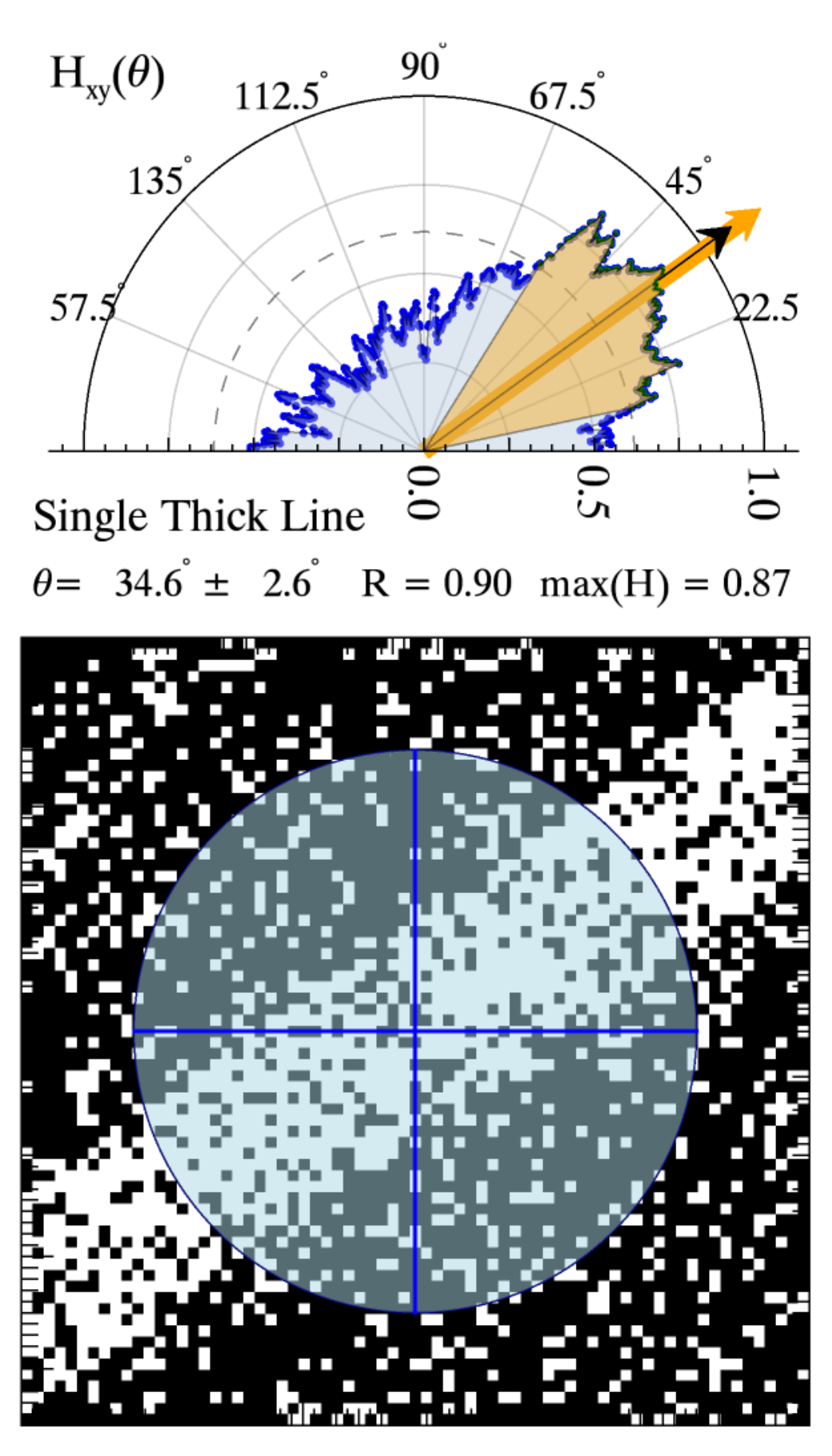} \includegraphics[width=0.31\textwidth]{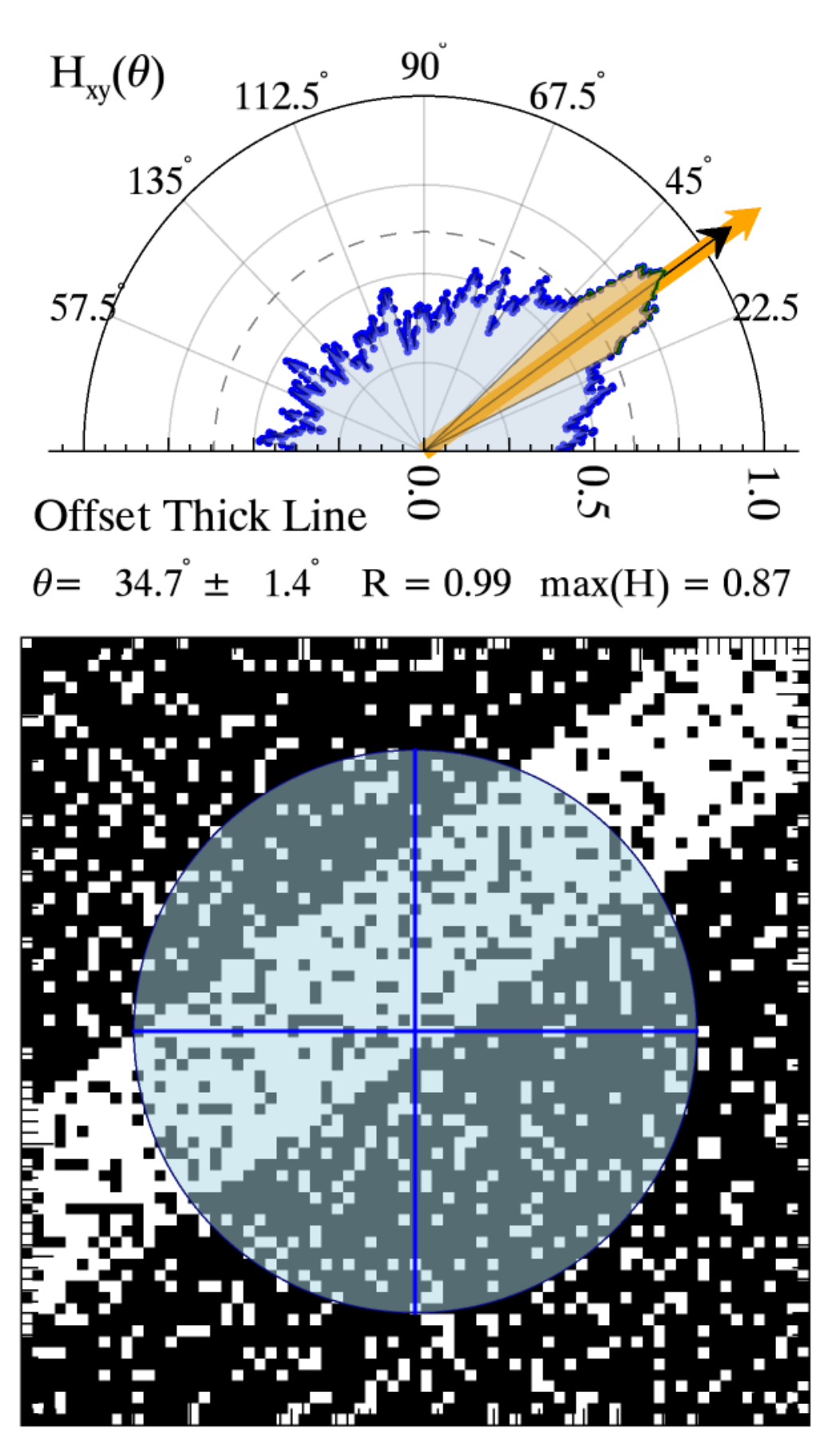} \includegraphics[width=0.31\textwidth]{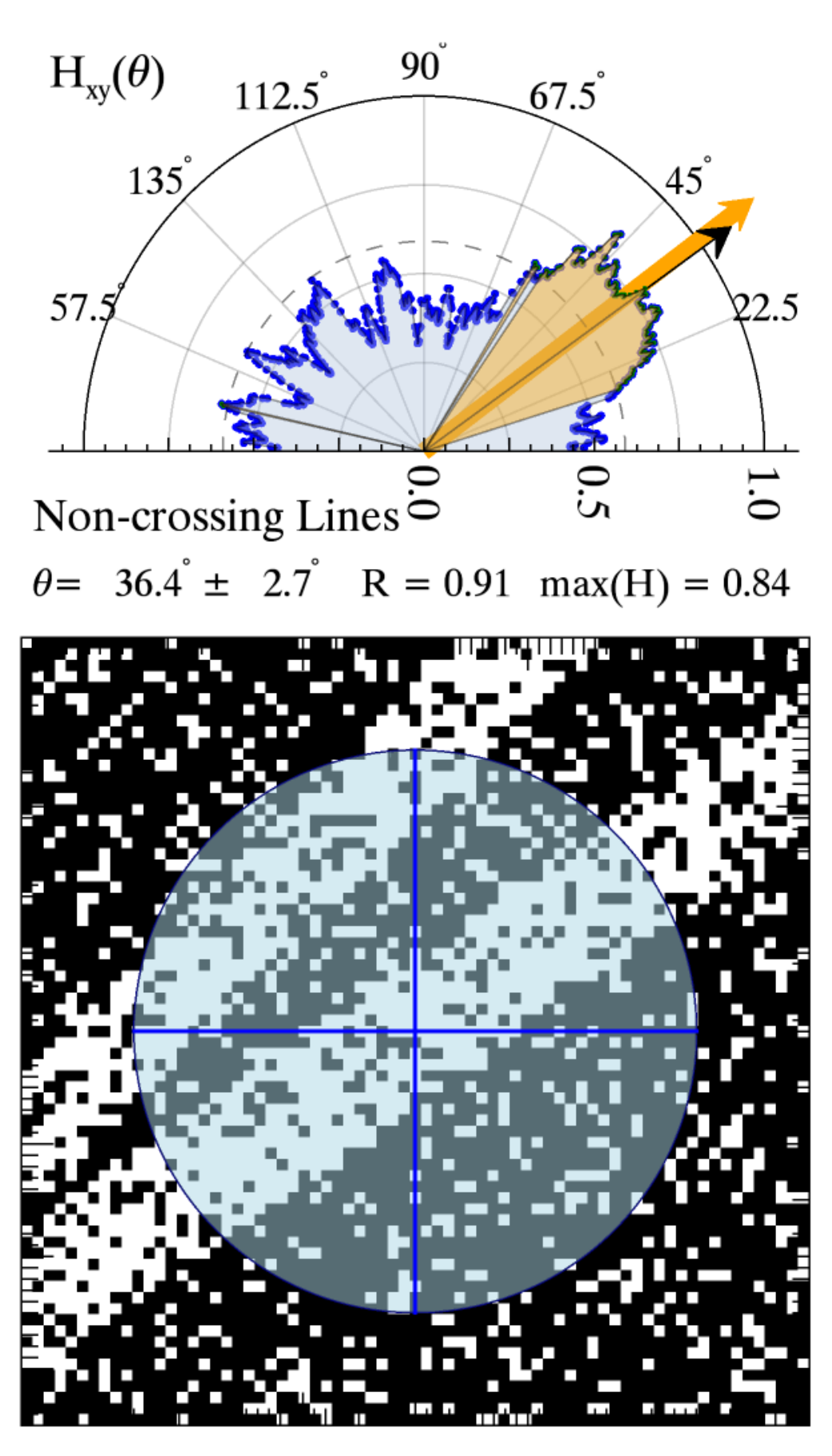} \\
\includegraphics[width=0.31\textwidth]{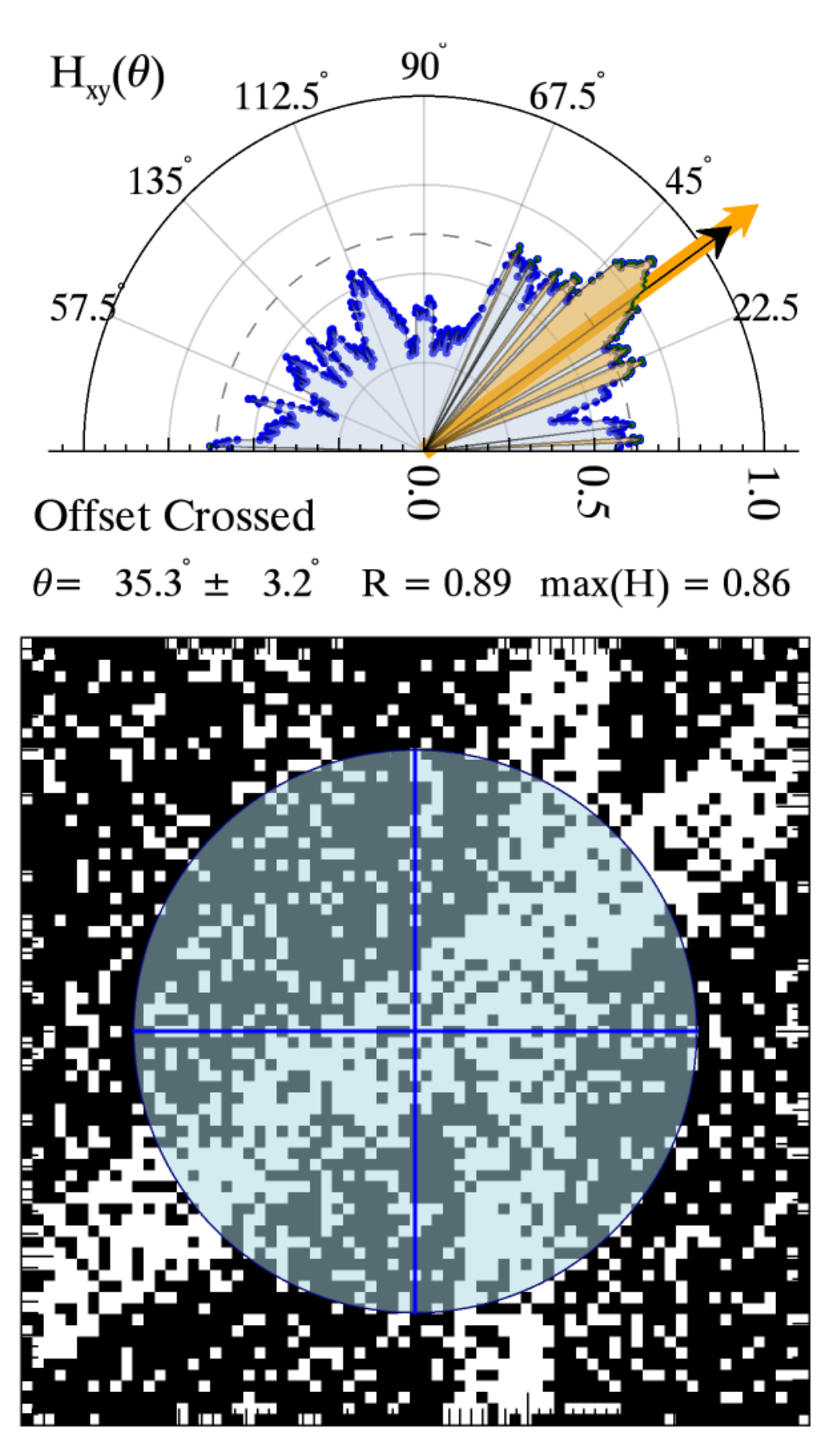} \includegraphics[width=0.31\textwidth]{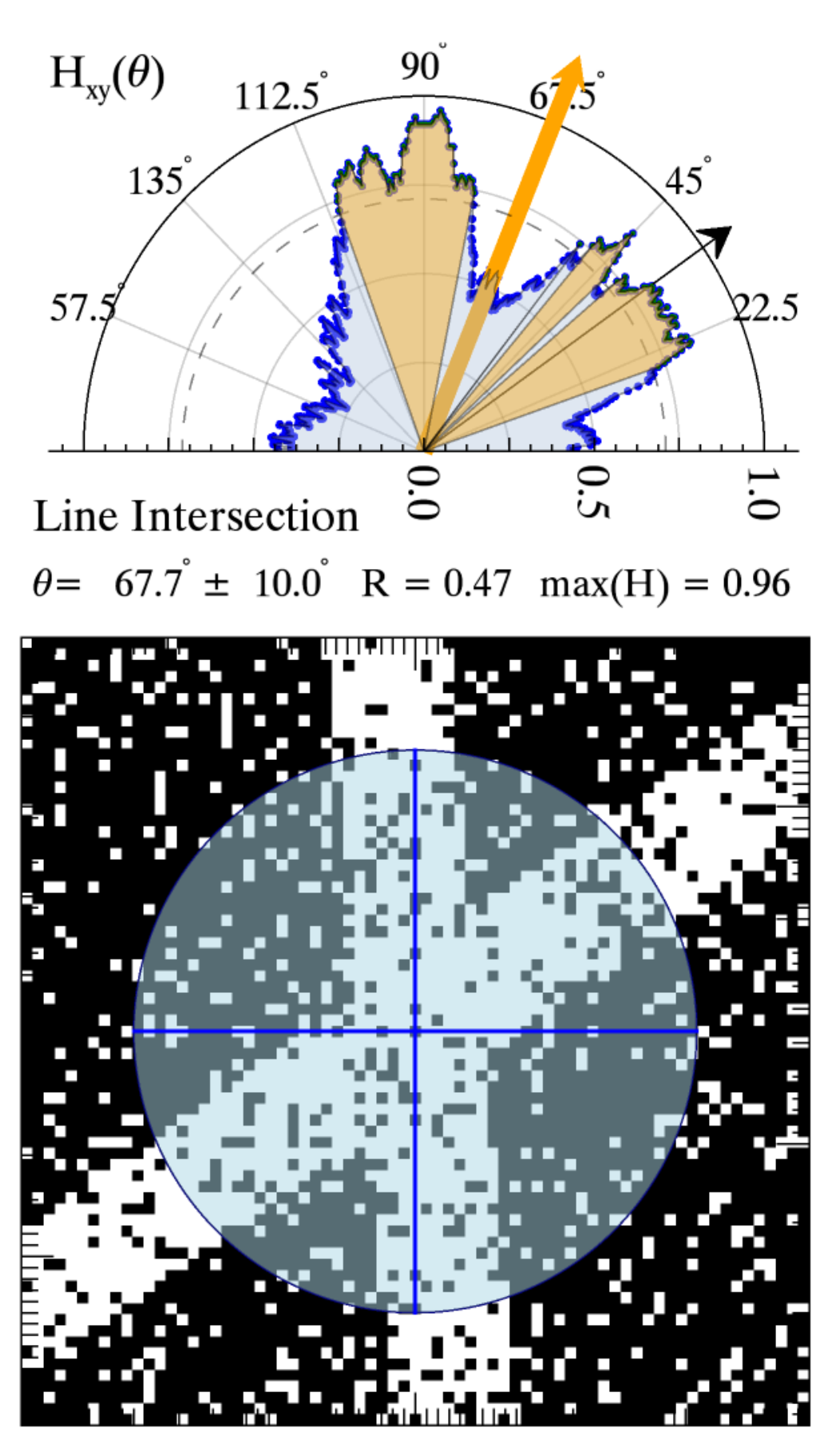} \includegraphics[width=0.31\textwidth]{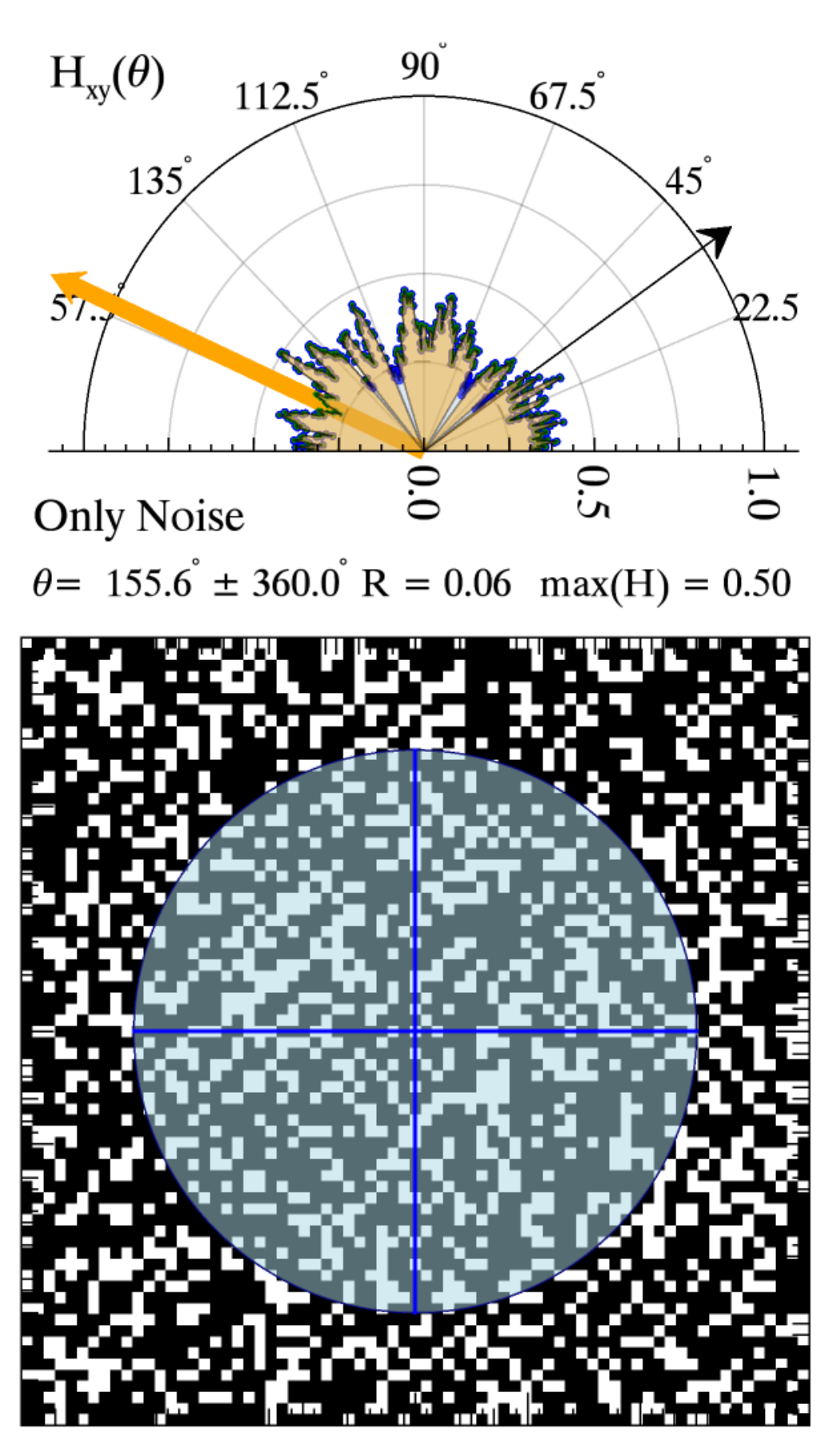} \\
\caption{Tests of RHT performance for common use cases encountered in solar imaging data sets using synthetic data with noise.  A line oriented at $35^{\circ}$ passes through the center of the image in each case except for the case of only noise.  Above each a polar plot of $H_{xy}(\theta)$ is displayed and the results from its calculation.  The adaptive threshold value is indicated by the dashed line, and every axis meeting the threshold is colored orange.  The black arrow is oriented at $35^{\circ}$ and the thick orange arrows points along the mean axial direction $\bar \theta$.}
\label{fig:example_use_cases}
\end{figure}

Illustrations of the above formalism for various use cases are given in Figure~\ref{fig:example_use_cases}.  Each image is a synthetic binary image with added noise containing lines of different thicknesses, relative orientations, and positions with respect to other lines and, importantly, the center of the circular kernel; that is, there is no assumption that the pixel is on a ridge.  The mean axial direction ($\bar\theta$), its 95\% confidence interval, the mean resultant length ($\bar R$) and the maximum of $H_{xy}(\theta)$ is reported in each case.  The RHT returns the true direction within the error bounds for all cases other than lines that intersect at the center of the circular kernel.  As shown by the offset thick line case, an advantage of the RHT is that it works for points not located on the ridge of the feature.  It is also not strongly affected by nearby lines that run nearly parallel or cross the feature away from the center of the kernel.  For the case of lines intersecting near the center of the kernel, $H_{xy}$ is a bimodal distribution and the returned mean axial direction is the average direction of the two lines (assuming similar thicknesses).  As this poorly represents the angle of either line, such cases need to be filtered from the RHT analysis of solar images,\footnote{It may be possible to fit a model to multimodal distributions such that intersecting lines may be individually identified, as was done for bird migratory patterns by \cite{ozarowska2013}.  Explicit fitting of the distribution, however, is numerically slow and not considered here.} which can be done based on the lower value of $\bar R$, \textit{i.e.} higher dispersion.  The same filtering removes points without sufficient linearity like shown in the noise only case in Figure~\ref{fig:example_use_cases}.  This ability to filter is a key advantage of this implementation of the RHT. 

\begin{figure}
\centering
\includegraphics[width=0.95\textwidth]{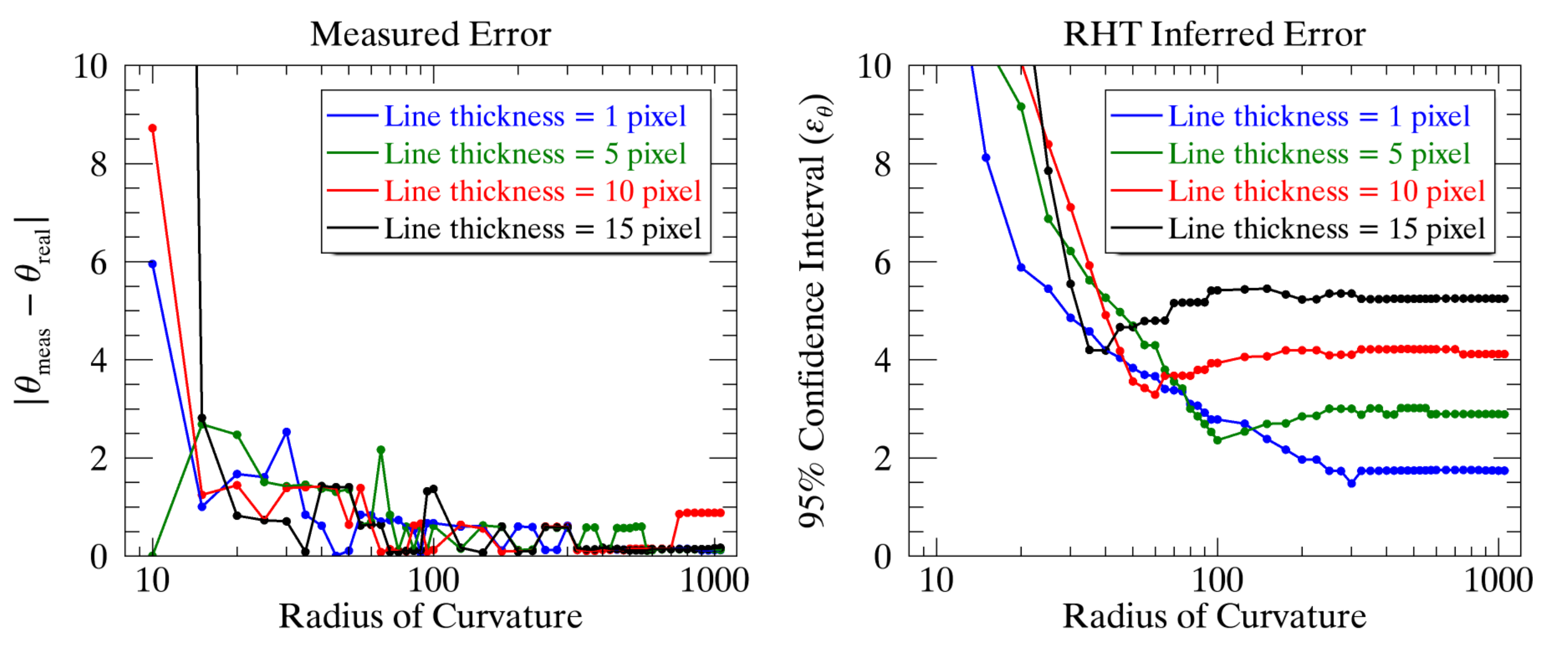}   
\caption{Tests of the RHT performance as a function of line thickness and line curvature for $D_{W} = 51$ pixels.  (\textit{left}) The measured angular difference between the mean axial direction derived from the RHT ($\theta_{meas}$) and the true angle ($\theta_{real} = 35^{\circ}$).  (\textit{right}) The 95\% confidence interval $\epsilon_{\theta}$ derived from the RHT using Equation~\ref{eqn:conf_int}.}
\label{fig:curvature_errors}
\end{figure}

To assess the errors introduced by the linear assumption for curvilinear lines, a simple numerical exercise was performed for lines of various thickness and radius of curvature and then subjected to the RHT analysis described above with $D_{W} = 51$ pixels.  The results are given in Figure~\ref{fig:curvature_errors}.  The error introduced by local curvature is not large for lines with uniform radii of curvature greater than $\sim20$ pixels.  In part this is due to the symmetry of $H_{xy}(\theta)$ as it samples a curved line; the curvature introduces extra dispersion in $H_{xy}(\theta)$ but the mean axial direction is less affected.  The 95\% confidence interval adequately characterizes the inherent error resulting from the linear approximation of a curved line.  The interval increases as the line thickness increases and the radius of curvature decreases.  It is also generally larger than the true error.    Therefore, the RHT approach is appropriate for the orientation analysis of curvilinear features that are not sharply curved. 

\subsection{Applying the RHT}  \label{sec:rht_howto}

The application of the RHT to solar imaging data requires the segmentation of the curvilinear features in the 2D spatial image and the creation of a binary image.  The full procedure is summarized as follows:
\begin{enumerate}[noitemsep]
\item Spatially filter the image to enhance the curvilinear features of interest.  Various filtering techniques may be applied including high-pass, band-pass, and/or edge filtering so long as features of interests, typically of some characteristic width, are enhanced relative to the background. 
\item Segment the features by binarizing the filtered image according to a specific threshold.  All pixels of interest, \textit{i.e.}, those associated with candidate curvilinear features to which the RHT is applied, need to take on a value of 1.  
\item Select parameters for the RHT.   The window width $D_{W}$ should be selected to be at least a few factors greater than the estimated average segmented feature width but kept small enough so that the features are mostly linear in the window.  The adaptive thresholding fraction $f$ may also be changed based on the level of noise in the image.  $f=0.25$ is typically a good value. 
\item Compute the RHT, $H_{xy}(\theta)$, for every pixel (x,y) in the binary image with value 1.  From $H_{xy}(\theta)$, compute $\max [ H_{xy}(\theta) ]$, $h_{xy}(\theta)$, $\bar \theta$, $\bar R$, and $\epsilon_{\theta}$. 
\item Filter results based upon $\max [H_{xy}(\theta) ]$, $\bar R$ and $\epsilon_{\theta}$. 
\end{enumerate}


\subsection{Application to coronal loop orientations} \label{sec:rht_trace}

To compare the RHT algorithm with other methods of tracing coronal loops, here it is applied to the problem of deriving the projected orientation of EUV coronal loops using the same coronal image used by \cite{aschwanden2008_compare} to compare five loop tracing codes. 


\subsubsection{Observations}

The observations, shown in Figure~\ref{fig:euv_results}(a), consist of a single EUV 171 $\AA$ image of NOAA active region 08222 observed by the \textit{Transition and Coronal Explorer} \citep[TRACE: ][]{handy1999} satellite on 19 May 1998 at 22:21 UT.  The image\footnote{Downloaded from \url{http://www.lmsal.com/~aschwand/software/tracing/tracing_tutorial1.html} on 9 June 2017.}, which have been processed with the standard TRACE\_PREP procedure, is a 1024 x 1024 pixel image, but, as in \cite{aschwanden2008_compare}, only the range $x =200-1000$ and $y=150-850$ pixels is used for analysis.  The spatial scale is $0.5''$ pixel$^{-1}$.


\subsubsection{Segmentation and RHT Parameters}

Spatial filtering is carried out in the manner of \cite{aschwanden2013} wherein a bandpass filtered image is created by the subtraction of two boxcar average smoothed versions of the EUV image.  To remain consistent, the boxcar width parameters of $n_{sm1} = 5$ and $n_{sm2}=7$ are used resulting in the image shown in Figure~\ref{fig:euv_results}(b).  A segmentation threshold of 0.5 (in units of data numbers), again to be compatible with the previous work, is used to binarize the data.  This threshold eliminates much of the interference fringe pattern apparent in the filtered image. The resulting binary image is shown in Figure~\ref{fig:euv_results}(c).  In this image, the majority of coronal loop widths are estimated to be in the range of 3 to 6 pixels, \textit{i.e.}, $1.5''$ to $3.0''$, and therefore an RHT kernel width of $D_{W} = 31$ pixels ($15.5''$) is selected for the RHT.  Small variations in the selection of $D_{W}$ do not significantly affect the results.  The window width is illustrated by the transparent blue circle is Figure~\ref{fig:euv_results}(c).


\subsubsection{Results}

\begin{figure}
\centering
\includegraphics[width=0.475\textwidth]{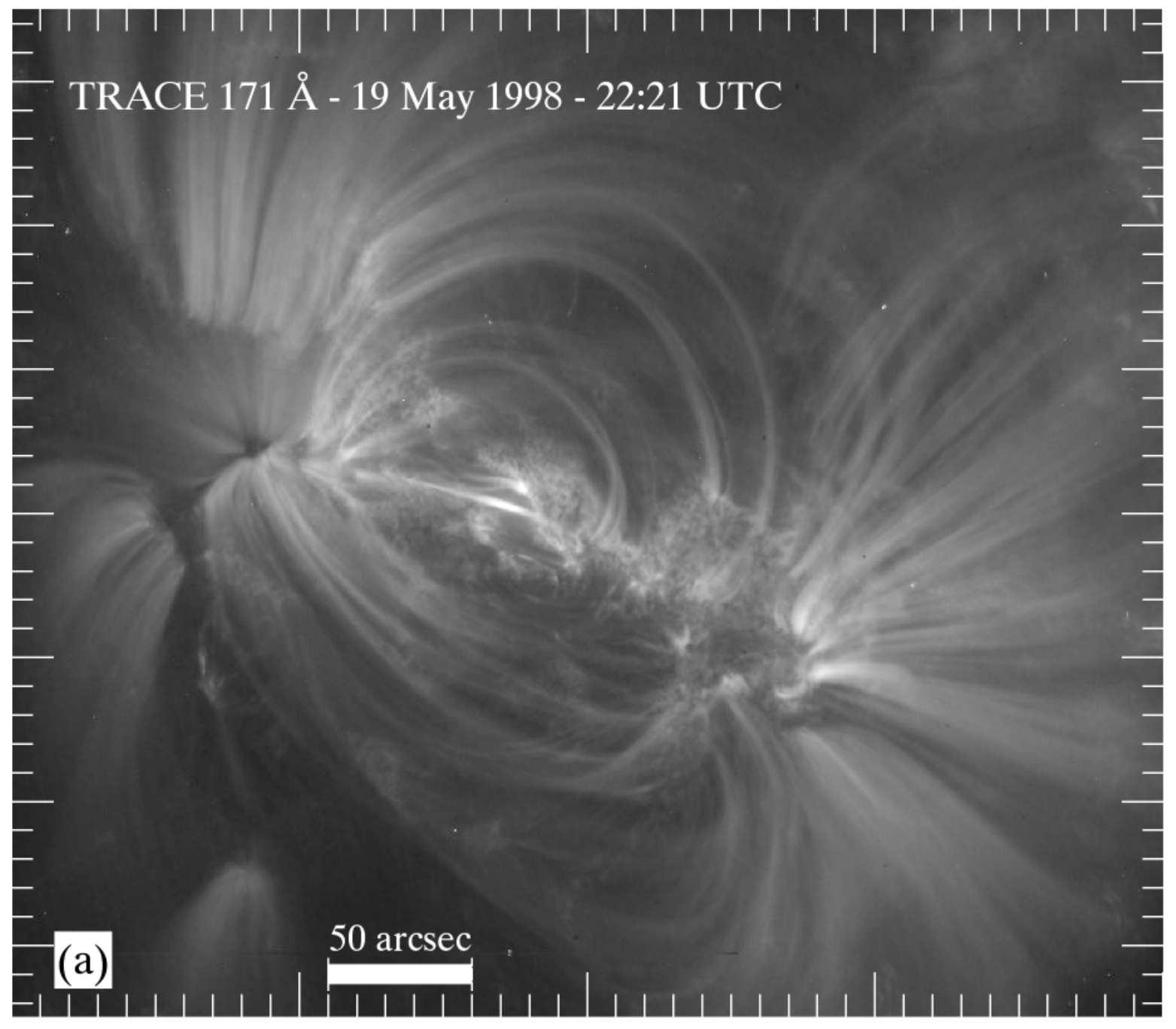} 
\includegraphics[width=0.475\textwidth]{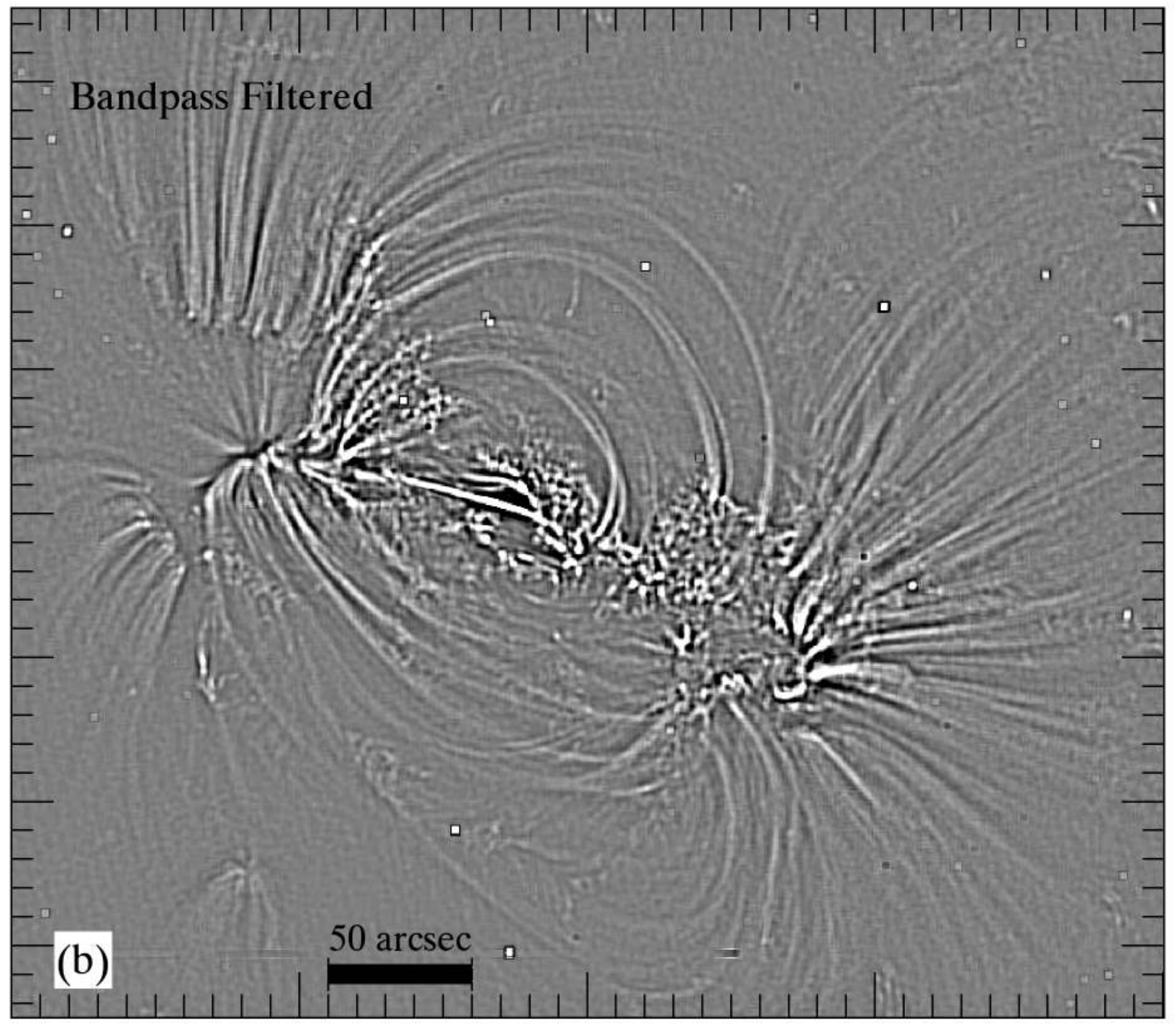} \\
\includegraphics[width=0.475\textwidth]{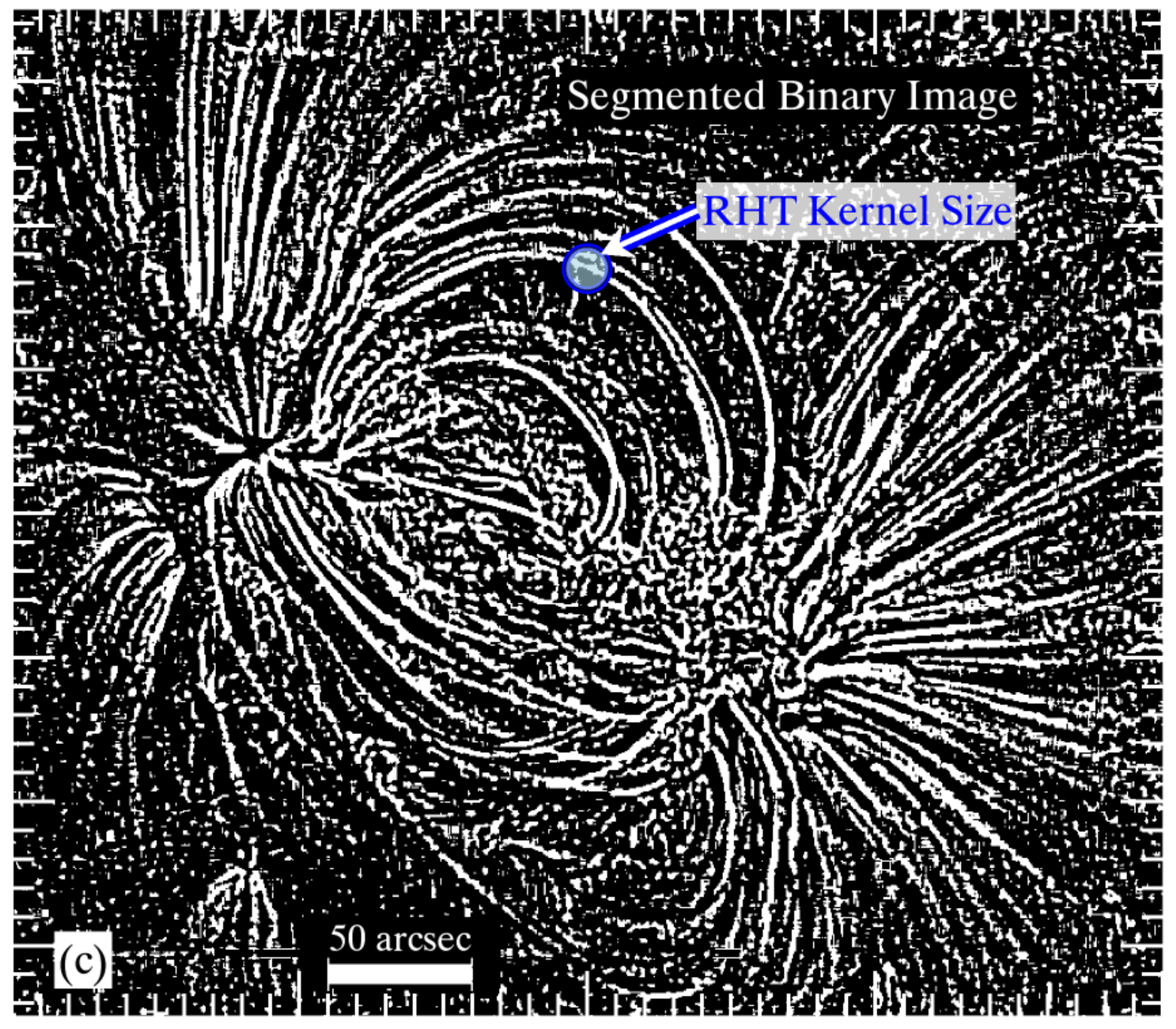} 
\includegraphics[width=0.475\textwidth]{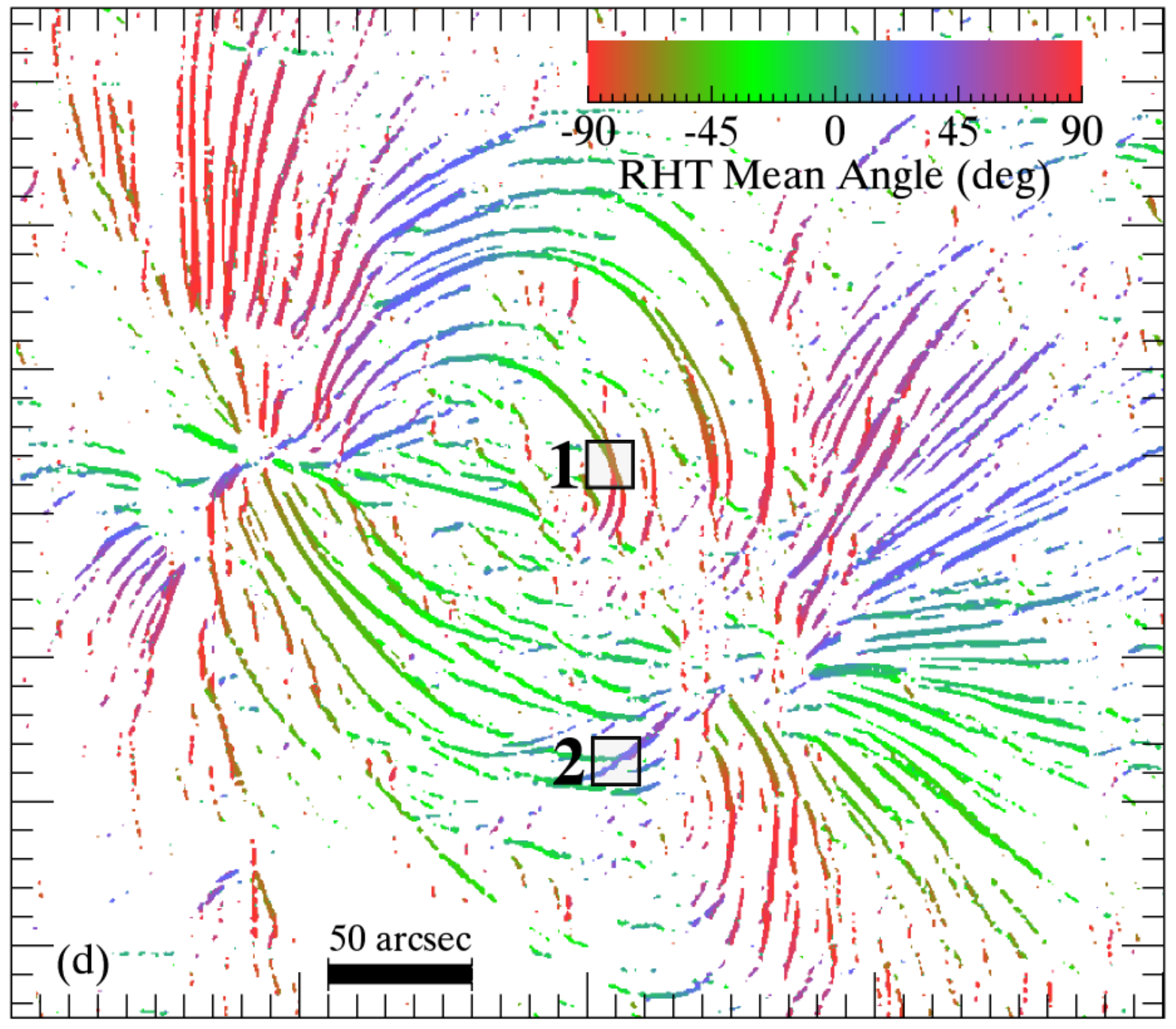} \\
\includegraphics[width=0.475\textwidth]{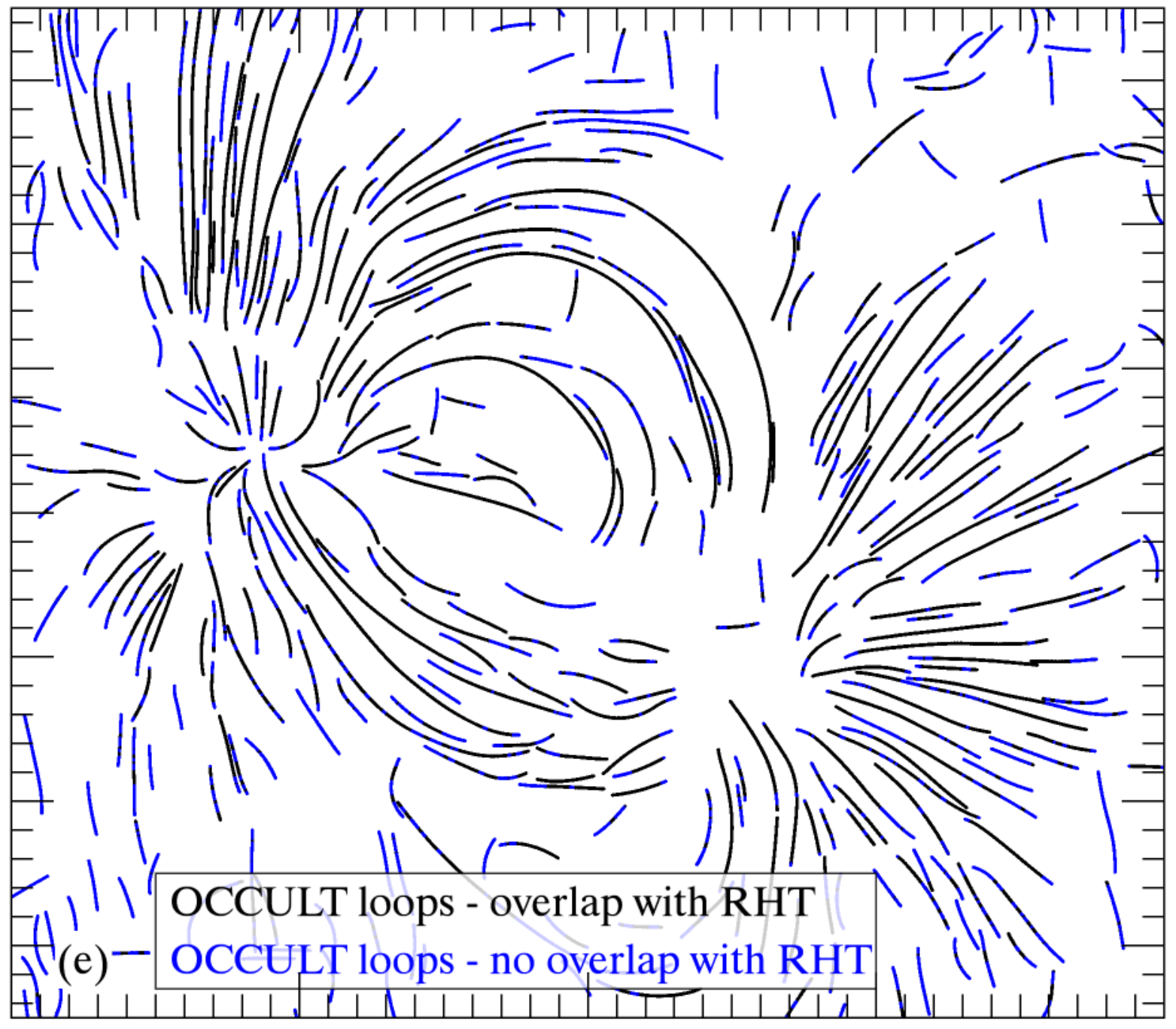} 
\includegraphics[width=0.475\textwidth]{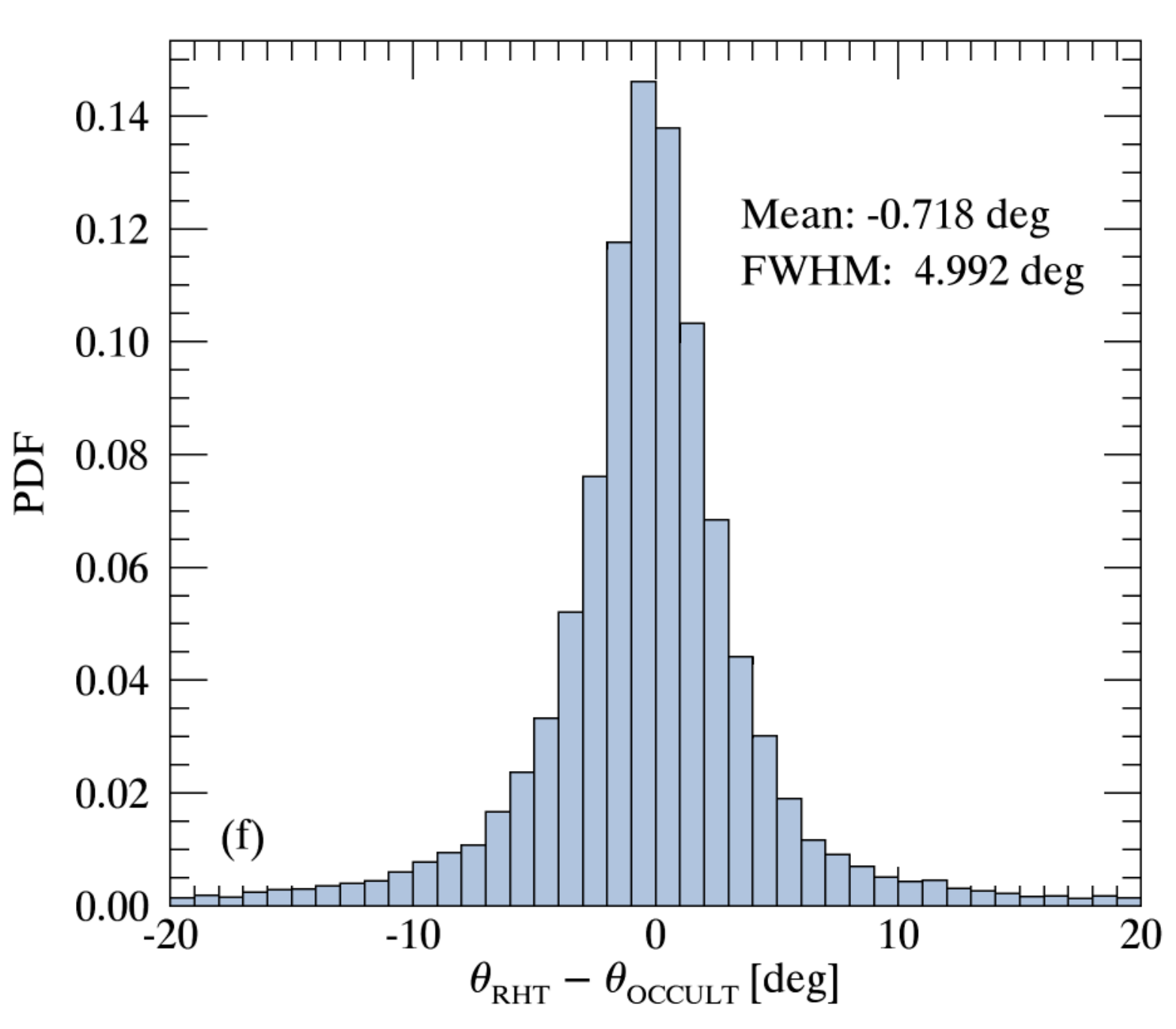} \\
\caption{Application of the RHT to derive EUV coronal loop orientations. (a) TRACE 171 \mbox{\AA} image of NOAA AR 08222 prior to processing. (b) Bandpass filtered version of the original image. (c) Binary image of the segmented coronal loop structures. Bright features correspond to the bright EUV coronal loop in panel a.  The kernel size used by the RHT analysis is indicated by the blue circle with a diameter of $(D_{W}-1) = 30$ pixels $(15'')$. (d) Mean orientaion angle results from the RHT procedure.  Boxes labeled `1' and `2' denote regions of intersecting features.  Angles increase in counterclockwise direction with a value of zero pointing to the right. (e) Map of the loop traces extracted by OCCULT where the color indicates whether the feature overlaps with the RHT results. (f) Histogram of the angular deviation between the OCCULT traces and the RHT derived orientation angles.}
\label{fig:euv_results}
\end{figure}

The mean projected orientations (azimuthal angles) derived from the applied RHT are displayed in Figure~\ref{fig:euv_results}(d) for all `1' valued pixels in the binarized image whose calculated mean resultant lengths ($\bar R$) are greater than 0.735 (corresponding to a circular standard deviation of $\sigma = 45^{\circ}$).  42\% of the $1.9 \times 10^{5}$ analyzed pixels from the binary image meet this criteria, with 28\% having $\bar R > 0.925$ ($\sigma = 18.5^{\circ}$).  The visible distribution of projected orientation angles for loops surrounding the loop footpoints is well represented by the RHT results as is the change in projected angle along individual loops connecting the two footpoints.  Importantly, the projected angle is well recovered for most pixels within each \textit{thick} segmented loop; that is, we do not rely on the presence of a ridge for axis definition and the results apply to a 2-D spatial structure, not a 1-D curve, as was illustrated in Section~\ref{sec:2DRHT} (Figure~\ref{fig:example_use_cases}).  

No ill edge effects are apparent at the outer boundary of the loop features, \textit{i.e.}, the derived orientation is consistent for all pixels within perpendicular slices along the loop.  Moreover, the presence of scattered disconnected areas does not signify poor RHT results.  Rather,  for many areas, only segments of the loop show sufficient directionality (above the image noise limit) for detection by the RHT.  The majority of these features have projected orientations consistent with nearby loop segments that do have noticeable lengths along their axes.  Due to the $\bar R$ based filtering, results negatively influenced by intersecting features are expected to be filtered out.  Cases where there is an intersection with reported results (see boxed areas in Figure~\ref{fig:euv_results}(d)) are features with shallow intersection angles (box 1) and where one feature has a greater thickness (box 2). 

To evaluate the results and compare with tracing-based methods of deriving the projected azimuth, the RHT results are compared with azimuths derived from OCCULT loop traces using procedures available in the SolarSoft library\footnote{See also \url{http://www.lmsal.com/~aschwand/software/tracing/tracing_tutorial1.html}}.  The version of OCCULT within \textit{looptracing\_auto4.pro} (last updated 8 December 2015) is used with the following input parameters:  $NSM1 = 5$, $RMIN = 30$, $LMIN=30$, $NSTRUC=10000$, $NLOOPMAX = 1000$, $NGAP = 1$, $THRESH1 = 0.50$, and $THRESH2= 1$.  See \cite{aschwanden2013} for parameter definition.  Approximately 67\% of the pixels in the resulting 604 traced loop segments shown in Figure~\ref{fig:euv_results}(e) have corresponding RHT-defined orientation values.  This fraction is not necessarily representative of performance but rather differences in curve detection near the noise limit.  The OCCULT method, using the above parameters, identies a number of interference fringes as loops that the RHT does not signify as having significant directionality on account of the applied threshold.  The traced loop projected orientation angles are derived from the local tangent of a 2nd-order polynomial fit in a least-squares manner to each point and its 4 closest neighbors along the loop after first spline interpolating the loop coordinates on an equidistant interval of 0.5 pixels.  A histogram of the angular separation between the RHT and OCCULT orientations, Figure~\ref{fig:euv_results}(f), provides evidence that the two algorithms determine consistent values for projected angles along loop ridges. 


\subsection{Application to chromospheric fine structure orientation} \label{sec:rht_ibis}

An advantage of the RHT is its ability to derive orientations for points located away from the ridge of the curvilinear features.  As was introduced above, the chromosphere's dense fine structuring provides a contrast map of orientation within a spatially-variant magnetic field that is likely much smoother than the visual impression created by the thermal fine structure.  As opposed to the coronal use case, the chromosphere is primarily optically thick thus lowering the expectation of visibly intersecting loops caused by the projection of a complex sparsely illuminated 3-D magnetic field.  As it is useful to have an estimate for the projected orientation of the chromosphere for pixels between ridges, here the RHT is tasked with generating a pseudo-continuous map of the chromosphere's azimuthal projection angle based on the fibril fine structure. 


\begin{figure}
\centering
\includegraphics[width=0.495\textwidth]{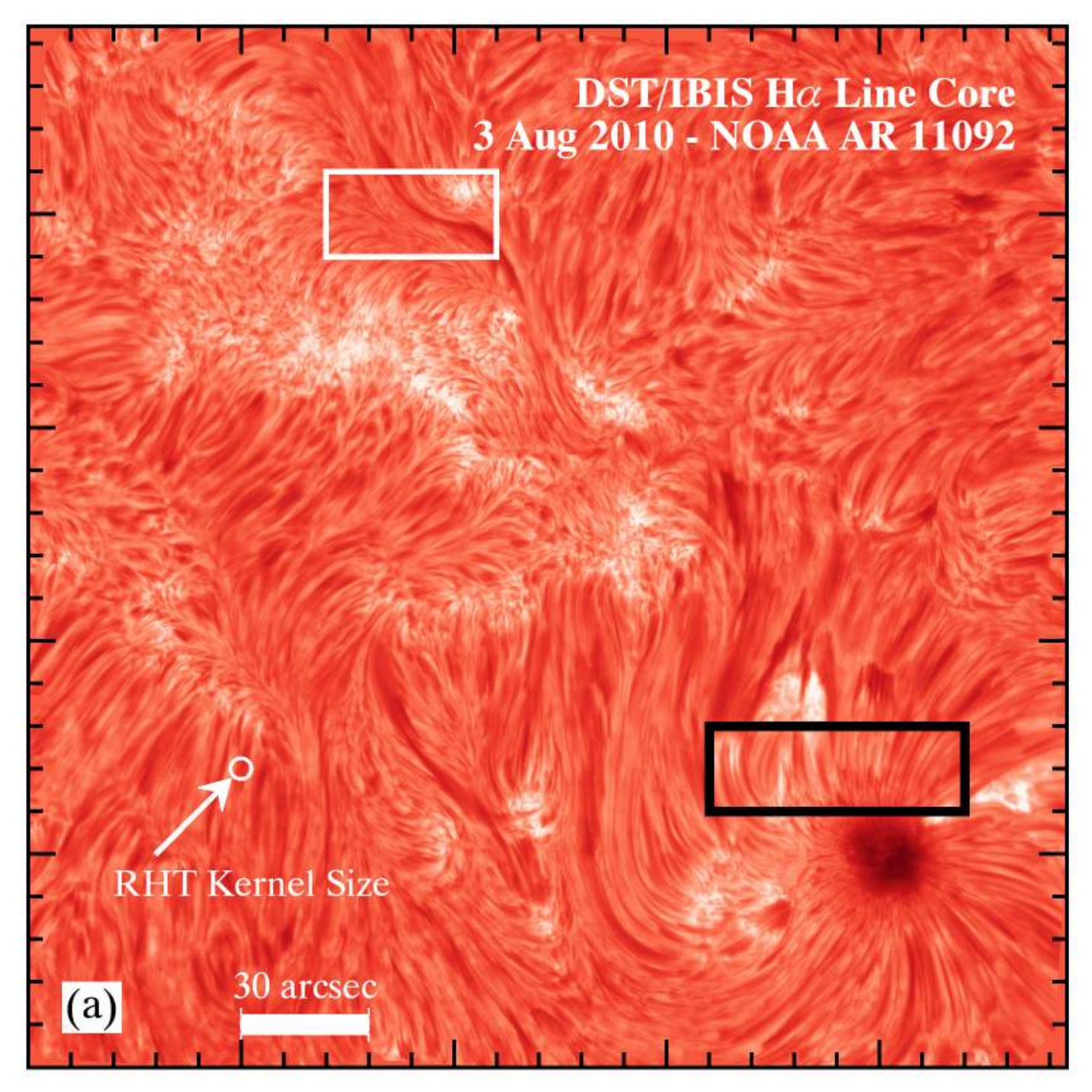} 
\includegraphics[width=0.495\textwidth]{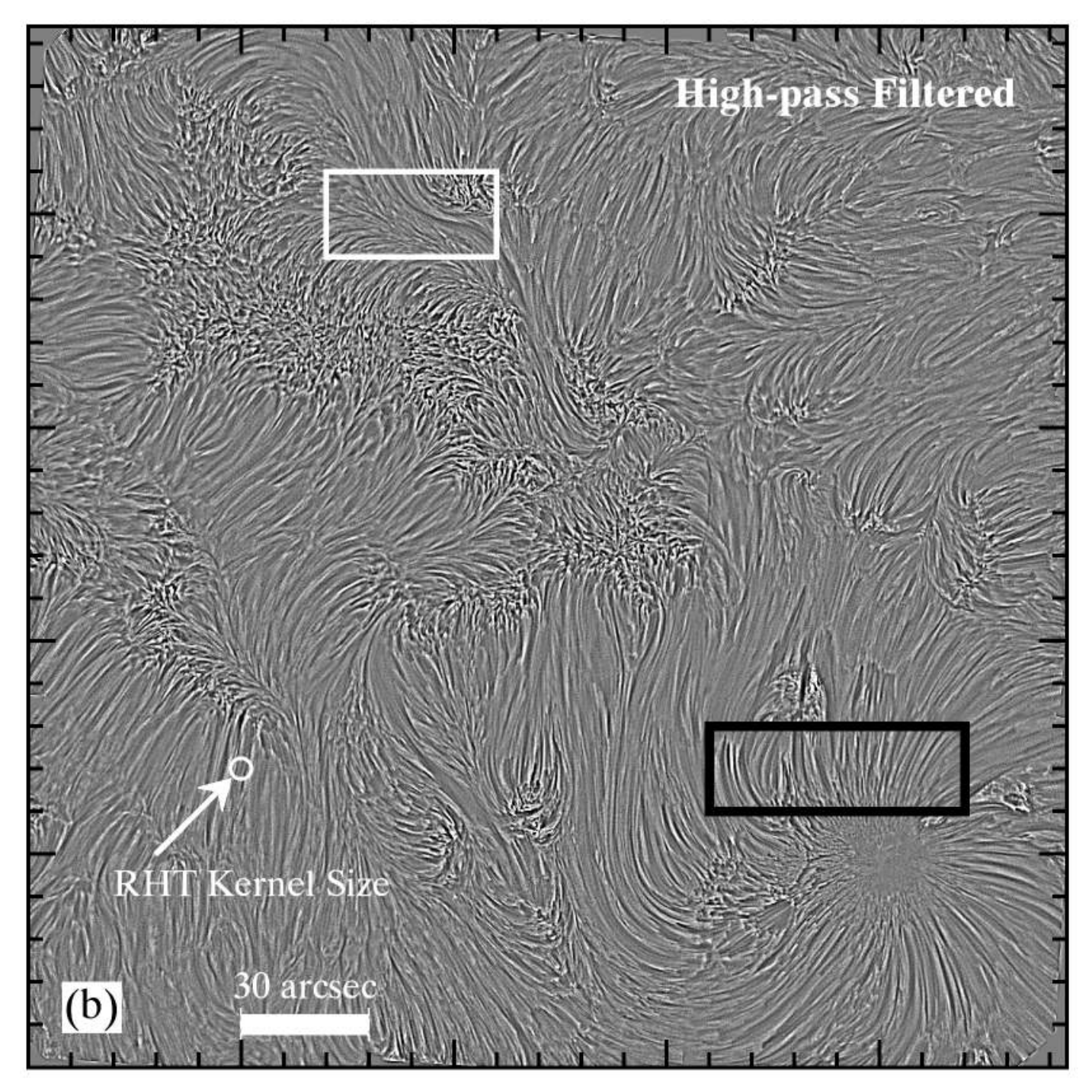} \\
\includegraphics[width=0.495\textwidth]{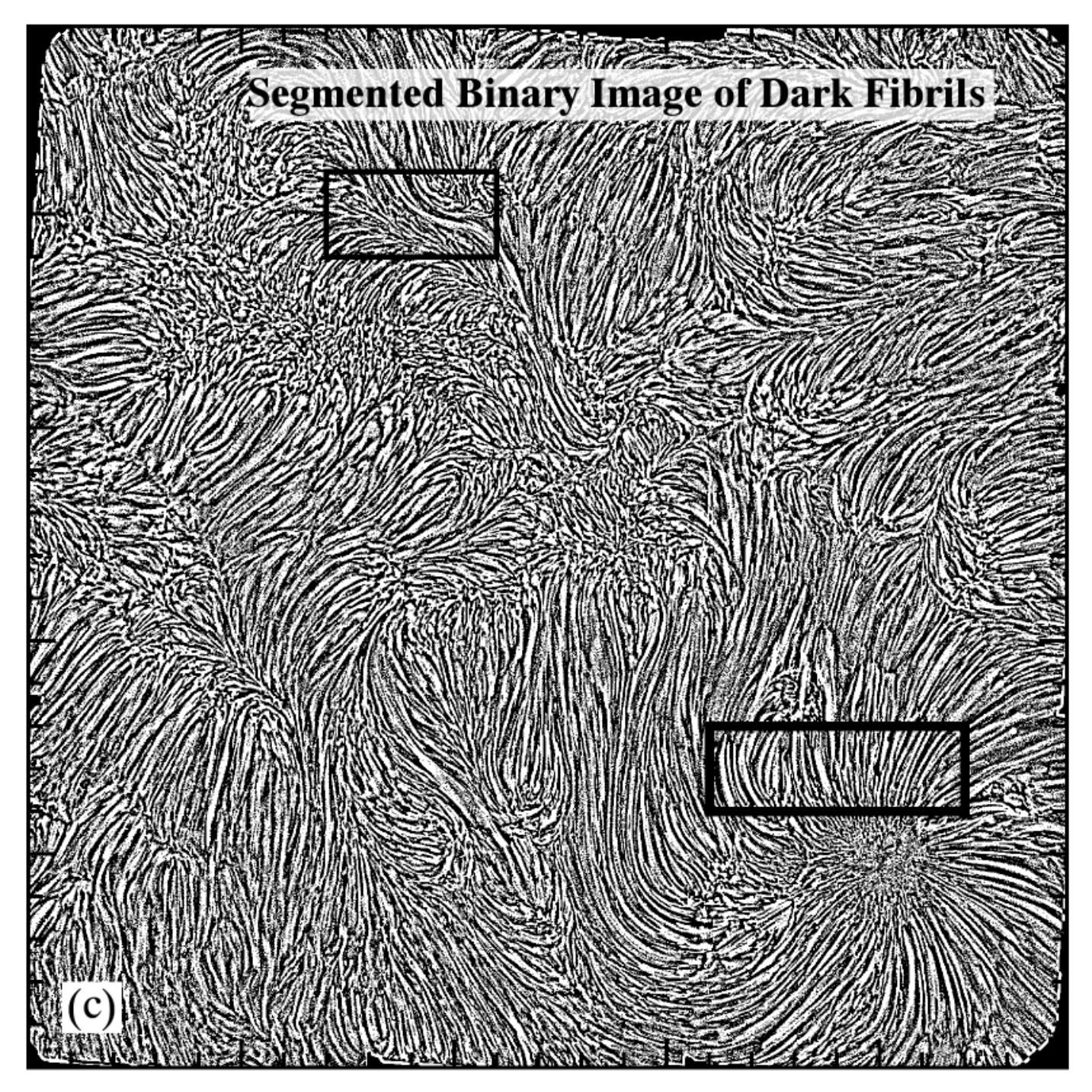} 
\includegraphics[width=0.495\textwidth]{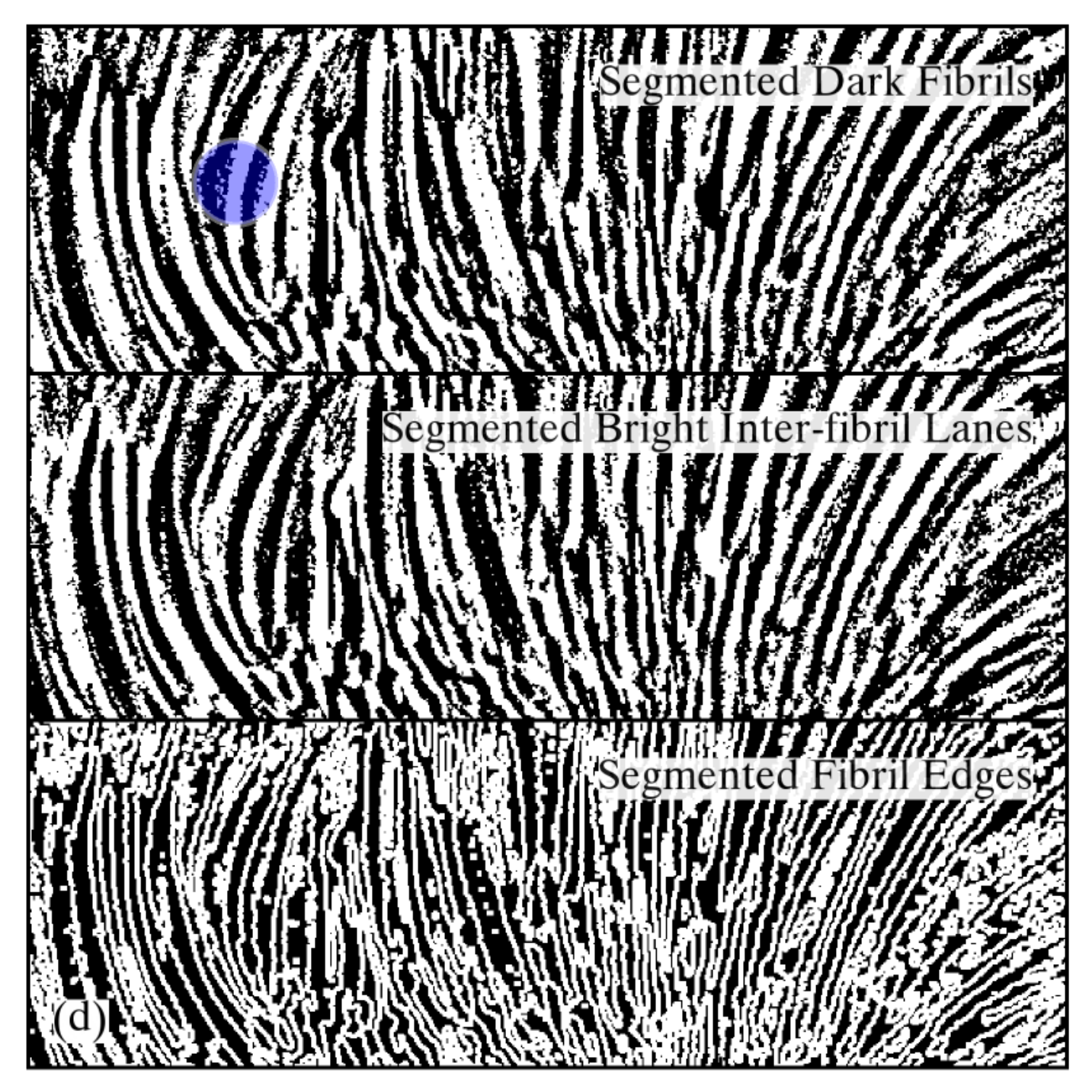} \\
\includegraphics[width=0.495\textwidth]{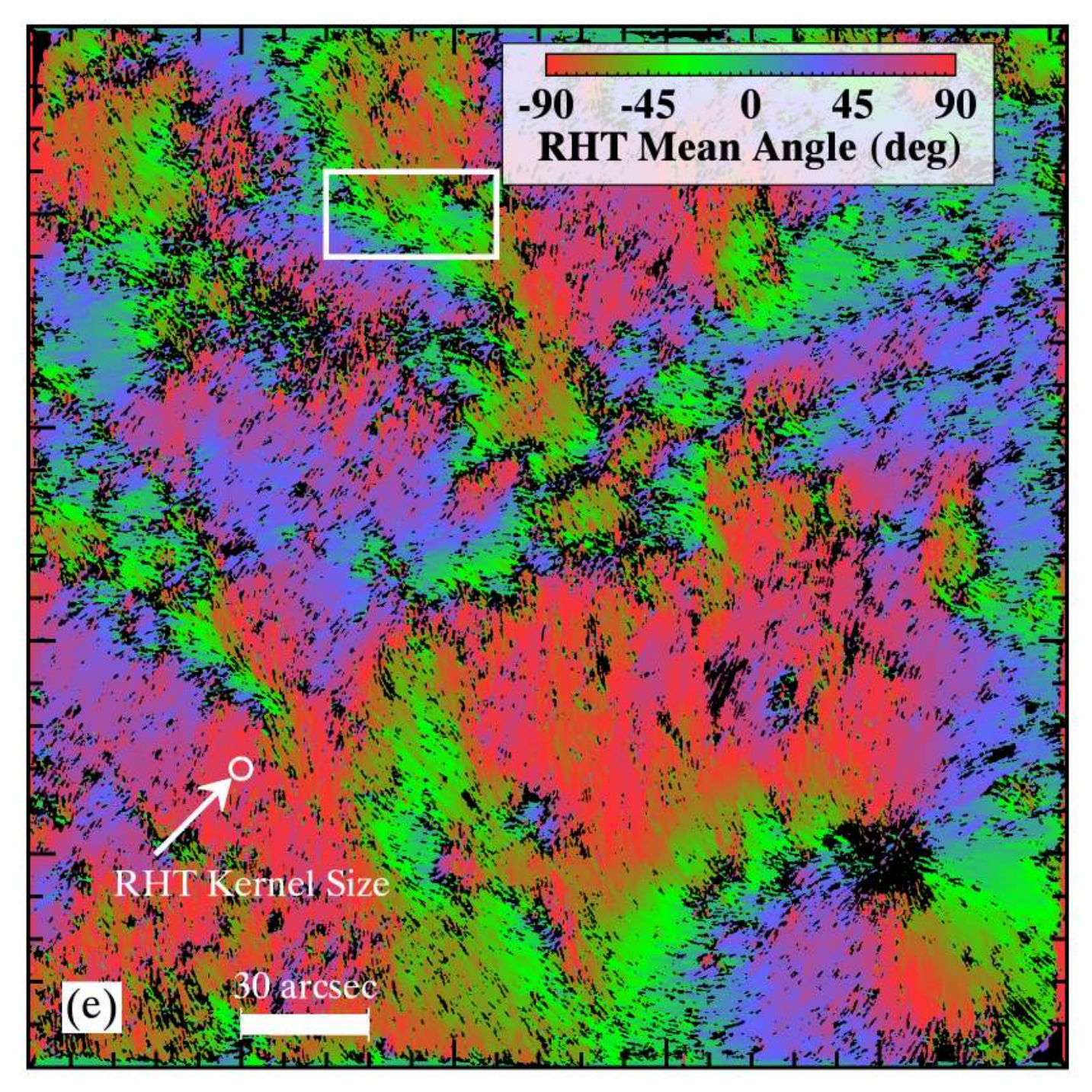} 
\includegraphics[width=0.495\textwidth]{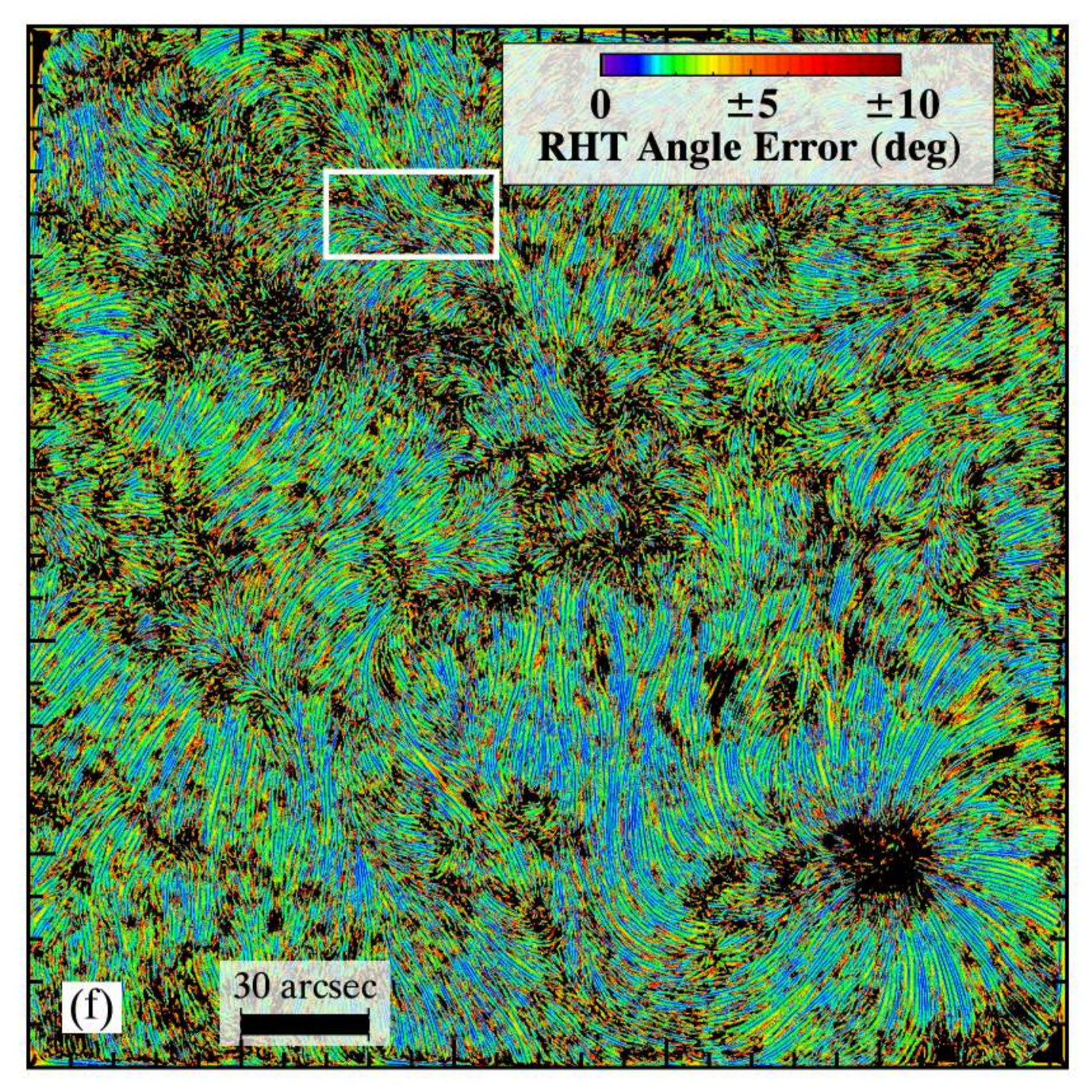} \\
\caption{Application of the RHT on chromospheric fibril orientations (a) IBIS H$\alpha$ image prior to processing.  (b) Ridge enhanced image segmented using the approach of Jing et al.  (c) Binary structure map of the fibrilar structure.  Bright features correspond to the darker, more absorbing, features in panel a.  The inverse of this image is processed by the RHT to derive orientation of bright regions.  (d) Close-up view of the three segmentation maps used for the quasi-continuous RHT procedure corresponding to the thick black rectangular in panel c. (e)  Results of the RHT algorithm showing the derived azimuthal angle.  Black regions correspond to those areas with mean resultant length less than 0.75 and normalized fraction less than Y.  (f) Error map for the derived azimuthal angle.}
\label{fig:ibis_rht_1}
\end{figure}


\subsubsection{Observations}

High spatial resolution H$\alpha$ observations of the chromospheric fine structure surrounding NOAA AR 11092 were obtained by the Interferometric BiDimensional Spectrometer \citep[IBIS:][]{cavallini2006,reardon2008} at the 76 cm aperture Dunn Solar Telescope on 3 August 2010.  IBIS is a dual Fabry-Perot based tunable narrowband filtergraph that iteratively scans through visible spectral lines.  The H$\alpha$ line core image analyzed here and shown in Figure~\ref{fig:ibis_rht_1}(a) has already been discussed in detail by \cite{cauzzi2012} and used for fibril detection by  \cite{jing2011} and \cite{aschwanden2016}.  Its large $244'' \times 244''$  field of view was obtained by a $3 \times 3$ stepped mosaic scan of the Sun across the instrument's $95''$ optical field of view.  The pixel size is $0.0976''$ square and the image quality has been improved through speckle deconvolution of narrowband image bursts ($1.22 \lambda / D$ @ $\lambda = 656.3$ nm is  $0.22 ''$). The region contains a large sunspot, superpenumbral fibrils, plage-related fibrils, an active region filament, and areas of spicular features.


\subsubsection{Segmentation and RHT parameters}

 \cite{jing2011} previously extracted direction information from the same data set using image segmentation, a union-finding algorithm to extract pixels corresponding to individual features, and polynomial fits to those pixels to derive the projected orientation angles along the features.  This procedure results in similar outputs as OCCULT, \textit{i.e.}, 1d coordinates of each identified feature, but does not necessary follow a particular feature ridge.  Here the segmentation approach of \citeauthor{jing2011} is adopted as the first step to the RHT analysis.  First, the image is highpass filtered by subtracting a Gaussian smoothed version of the image with $\sigma = 5$ and a kernel width of $15 \times 15$ pixels.  The threshold used to binarize the image is defined by \citeauthor{jing2011} as $\mu$ times the median of the highpass filtered image where $\mu$ is manually selected to be $7/8$.  Pixels with values below this threshold are set to 1 and otherwise 0.  Figure~\ref{fig:ibis_rht_1}(b) and (c) show the highpass filtered image and the binarized image, respectively.
 
To create a pseudo-continuous map of projected orientation, three binary images are provided as input to the RHT algorithm.  The first is that defined above which isolates the dark features in the original image.  The second binary image is an inverted version of the first, which identifies the bright areas between the dark features.  Finally, since the edges of features in these maps are not sharply defined, the peak of the RHT for these pixels may struggle to be significantly higher than the contribution of the surrounding pixels.  Thus, as a third binary image, we create an edge-filtered version of the two previous binary maps by applying a Sobel edge filter and then dilating the results with a $2 \times 2$ boxcar kernel (see Figure~\ref{fig:ibis_rht_1}(d)). For pixels that exist in multiple binary images, the results with the highest $\bar R$ are selected. 

As most feature widths are between 5 and 10 pixels ($0.49'' - 0.98''$), the RHT kernel width is set to $D_{W} = 51$ pixels ($4.98''$) as indicated in Figure~\ref{fig:ibis_rht_1}. Small variations in the selection of $D_{W}$ do not significantly affect the results.


\subsubsection{Results}\label{sec:ibis_results}

\begin{figure}
\centering
\includegraphics[width=0.495\textwidth]{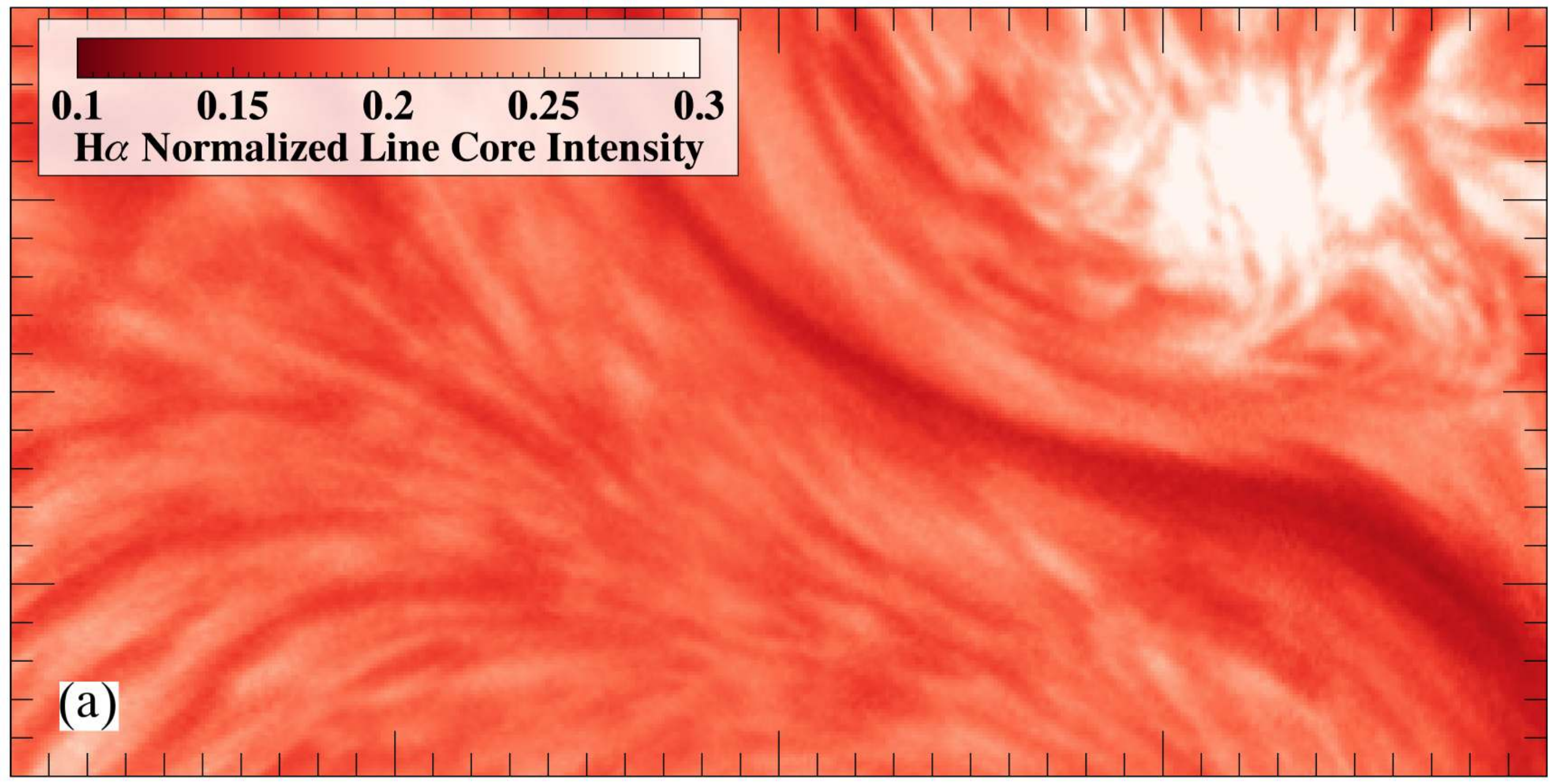} 
\includegraphics[width=0.495\textwidth]{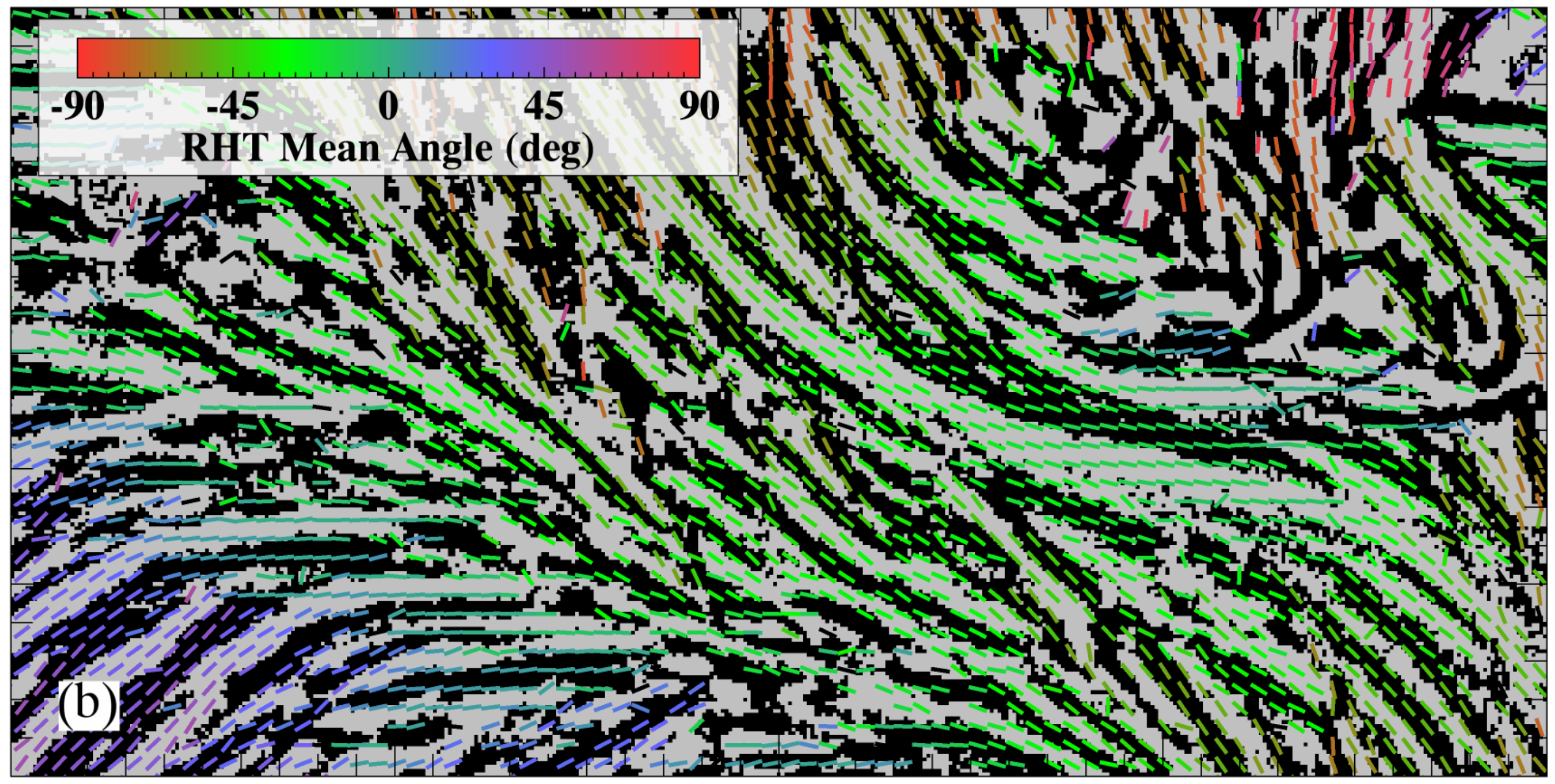} \\
\includegraphics[width=0.495\textwidth]{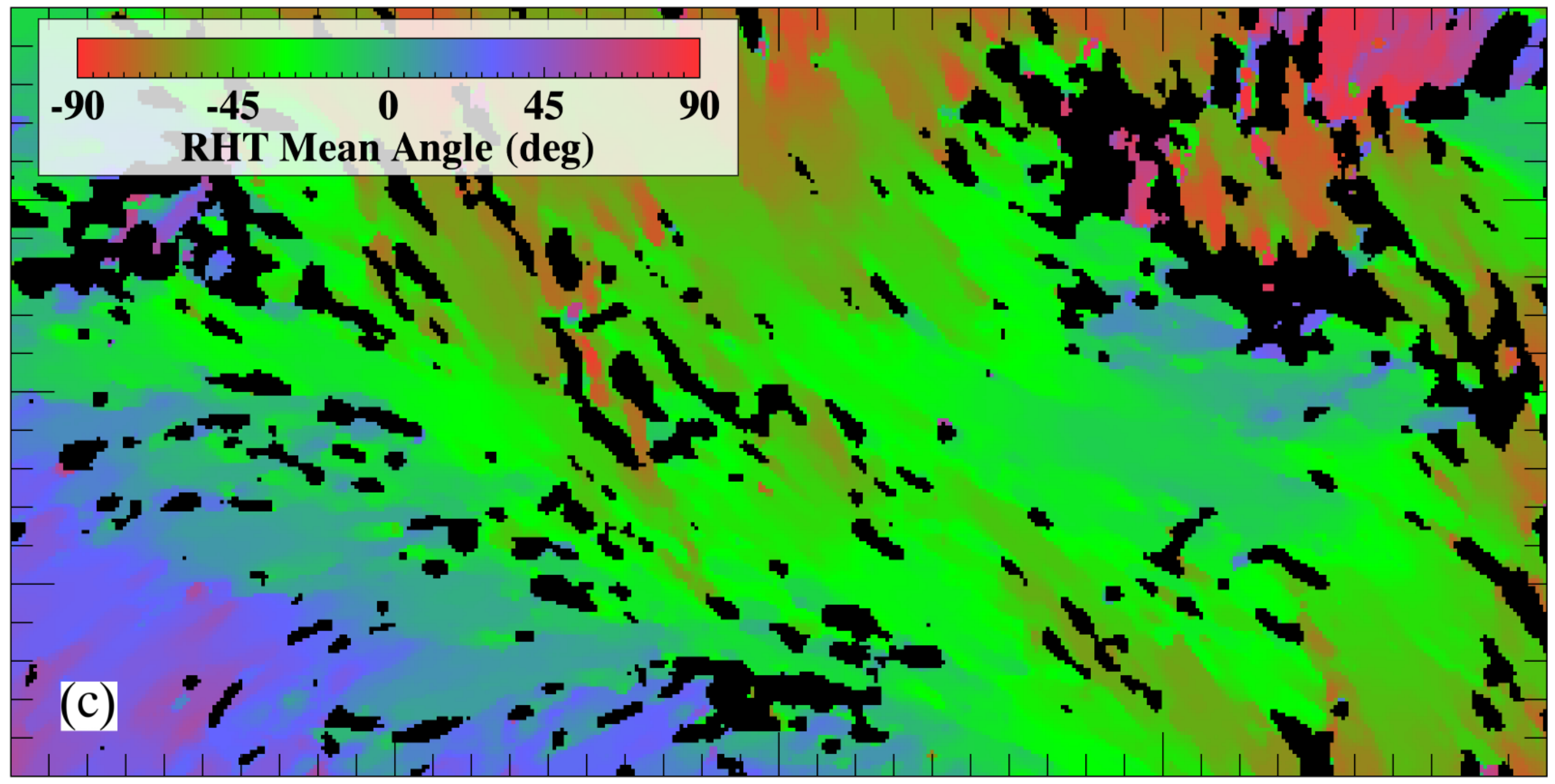} 
\includegraphics[width=0.495\textwidth]{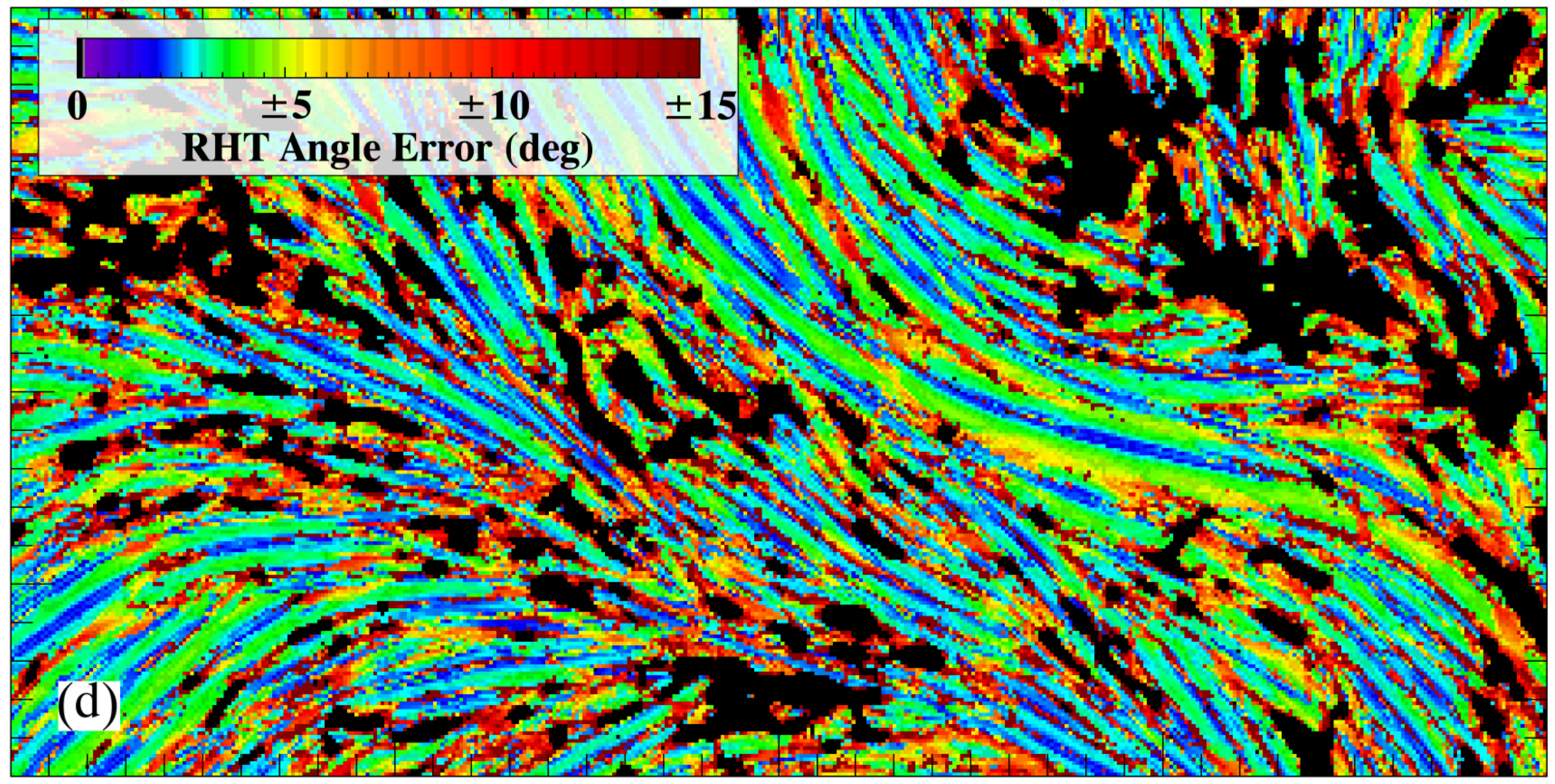} \\
\caption{Magnified region of interest of the chromospheric H$\alpha$ line core image (a) and the RHT results as identified in Figure~\ref{fig:ibis_rht_1}.  Panel (b) is a vector plot of the the RHT mean axial directions overplotted on the binary map of the dark chromospheric fine structure (displayed with reduced contrast to enhance visibly of the vectors).   All pixels with $\bar R < 0.735$ are colored black in panels (c) and (d) and are not included in the vector plot of panel (b).}
\label{fig:ibis_rht_2}
\end{figure}

Figures~\ref{fig:ibis_rht_1} (panels e and f) and~\ref{fig:ibis_rht_2} illustrate the results of the RHT as applied to the IBIS chromospheric data set.  $82\%$ of the $2500^2$ image pixels have $\bar R > 0.735$ ($59\%$ with $\bar R > 0.925$).  The $18\%$ of the pixels with $\bar R < 0.735$ are colored black in the figures and mostly cluster in regions above plage where the chromosphere exhibits less distinctive linear coherence.  Outside of these regions, the process succeeds in generating a pseudo-continuous map of the chromospheric fine structure orientation and the orientations show spatial variation visibly consistent with the fine structure orientation in the original image.  The $95\%$ confidence intervals for pixels of $\bar R > 0.735$ are in general less than $\pm 10^{\circ}$ and less than $\pm 5^{\circ}$ along the structures in the binarized edge map, which, due to the thin character of an edge, lead to a smaller RHT error than a thick feature (for the same kernel size).

A vector plot of the RHT mean directions (with $\bar R >0.735$) overploted on a magnified region of interest (ROI) is displayed in Figure~\ref{fig:ibis_rht_2}.  The RHT mean directions and the feature orientations are consistent for both white (\textit{i.e.}, corresponding to dark features in the original image) and black regions in the binary map.  In the bright plage region on the upper right of the ROI, the majority of pixels do not return significant RHT mean directions.  However, there are areas in the center that show orientation, visibly in the binary maps, as well as in the RHT results.  While linear coherence exists here, features like these must be analyzed with caution as linearity may not be emblematic of, \textit{e.g.}, a fibril associated with a field orientation, but rather originate from some other macrostructuring process.  

\section{RHT-Assisted Automated Time-Slice Motion Analysis}


\subsection{Hierarchical multi-dimensional application of the RHT}  \label{sec:time_slice_rht}

Having applied the RHT to two-dimensional curvilinear spatial structures, it is natural to extend the method to automated time-slice analysis of apparent motion for use cases where that motion traces out three-dimensional curvilinear paths in imaging time-series data.  The problem amounts to quantifying the coherence and directionality of a 3D structure.  Such line detection methods based on the original Hough transform technique have been advanced for 3D geometries, \textit{e.g.}, by \cite{jeltsch2016}.  For time-slice analysis of apparent motion, these approaches can be adapted and further simplified by considering only the line \textit{direction} parameters in a spherical geometry, \textit{i.e.,} azimuth and elevation, on a point-by-point basis, \textit{i.e.}, in the same manner as the 2D rolling approach by \clark.  The two direction parameters correspond to the projected angle of motion in the plane of the sky and the material's velocity.  Synthetic tests of a 3D RHT method for oriented point clouds using the spherical discretization approach of \citeauthor{jeltsch2016} for the accumulator, \textit{i.e.,} vertex directions of tesselated platonic solids, did successfully extract known orientation angles.  However, in turning to the application to real time-slice motion analysis, the 3D RHT approach proved less than optimal only because the trajectories of the apparent motion could be better segmented by applying separate filtering techniques for the spatial and temporal domains. 

In lieu of using a 3D version of the RHT, a hierarchical approach based on the 2D RHT is developed here for automated time-slice analysis that allows one to apply different segmentation techniques for each domain.  The algorithm first finds the projected orientation of the apparent flow based on a 2D RHT analysis of each individual frame after a running temporal mean and spatial segmentation procedure is applied.  The running mean is required so that the material's projected angle along its flow axis is discernible in a single frame.  Based on those results, the apparent projected velocity is then derived by applying the 2D RHT to a time-slice oriented at the projected angle derived in the first step, in a manner similar to how manual time-slice analysis is routinely performed.  The full procedure can be summarized as follows: 
\begin{enumerate}[noitemsep]
\item Calculate running (simple moving) temporal mean of imaging time series over a centered kernel of width $w_{r}$ such that the projected angle of features with apparent motion is discernible in a single frame over a path length a few factors larger than the feature width. 
\item Spatially filter each resulting frame to enhance the apparent curvilinear features, which may correspond to static features or features with apparent motion.  Once again, various filtering techniques may be applied. 
\item Segment the features in each spatially filtered frame by binarizing according to a specified threshold.  All pixels of interest, \textit{i.e.}, those for which the RHT is applied, should have a value of 1.  
\item Return to the original data and apply a temporal filter to enhance the pixels through which the apparent motion trajectories traverse.  Various temporal filters may be applied provided no phase lag is introduced.
\item Segment the temporally filtered data by binarizing according to a specified threshold.  All pixels of interest, \textit{i.e.}, those for which the RHT is applied, should have a value of 1. 
\item Select parameters for the RHT.  The window width in the spatial domain $D_{W}|_{xy}$ should be selected, as before, based on the average feature width. The window width in the (spatio-)temporal domain $D_{W}|_{t}$ can similarly be selected by feature width and may be different than $D_{W}|_{s}$; although, here the two widths are kept equal $D_{W} = D_{W}|_{xy} = D_{W}|_{t}$.  The adaptive thresholding fraction $f$ may also be individually specified though is typically kept at $f=0.25$ for the use cases here. 
\item For each frame in the imaging time series, perform the following:
\begin{itemize}
\item Compute the spatial 2D RHT function, $H_{xy}(\theta)$, for every pixel (x,y) in the spatially segmented image with value 1.  From $H_{xy}(\theta)$, compute $\max [ H_{xy}(\theta) ]$, $h_{xy}(\theta)$, $\bar\theta_{xy}$, $\bar R_{xy}$, and $\epsilon_{\theta}|_{xy}$. 
\item For every analyzed pixel from previous step whose $\bar R_{xy}$ is above some criteria (typically $\bar R_{xy} > 0.75$), extract the pixels from the temporally segmented data cube in the time-slice oriented at $\bar\theta_{xy}$ and within the (spatio-)temporal kernel width $D_{W}|_{t}$. Then compute the temporal 2D RHT, $H_{t}(\theta)$, and derive $\max [ H_{t}(\theta) ]$, $h_{t}(\theta)$, $\bar\theta_{t}$, $\bar R_{t}$, and $\epsilon_{\theta}|_{t}$. 
\end{itemize}
\item Filter results based upon $\max [H_{xy}(\theta)]$, $\bar R_{xy}$, $\epsilon_{\theta}|_{xy}$, $\max [H_{t}(\theta)]$, $\bar R_{t}$, and $\epsilon_{\theta}|_{t}$. 
\item Compute components of projected apparent velocity using $\bar\theta_{xy}$ and $\bar\theta_{t}$, as shown below. 
\end{enumerate}


\subsection{Application to coronal rain dynamics} \label{sec:rht_iris}

The hierarchical RHT-assisted apparent motion analysis is demonstrated here for coronal rain observed off-limb by the Interface Region Imaging Spectrograph \citep[IRIS:][]{depontieu2014}. 


\subsubsection{Observations}

Observations of NOAA Active Region 12468 were obtained near the east solar limb (centered at $X=-1017'';Y=-209''$) by IRIS on 9 December 2015.  IRIS carried out OBS ID 3620259404 consisting of a very large sit-and-stare observation wth the maximum field of view of $167'' \times 174''$.  Both 1400 \mbox{\AA} and 2796 \mbox{\AA} slit jaw imager (SJI) observations were obtained at a 19 second cadence between 17:41 and 18:40 UT using an exposure time of 8 seconds.  Only the 1400 \mbox{\AA} observations, which are dominated by the Si \textsc{iv} transition region lines at $\lambda1393.78$ and $\lambda1402.77$ (both formed near $\log T \approx 4.8$ K), are used here.  The IRIS/SJI image scale is $0.166''$ pixel$^{-1}$ with a spatial resolution $0.33''$ for the 1400 \mbox{\AA} observations. 

The Level 2 data products are used for analysis; however, an additional correction for instrument wobble was made by cross correlating features near the limb and applying shifts to the images to compensate.  In addition, the world coordinate system (WCS) pointing information is refined by coaligning an individual 1400 \mbox{\AA} SJI image with a co-temporal 1700 \mbox{\AA} full-disk image acquired by NASA's Solar Dynamics Observatory's Atmospheric Imaging Assembly \citep[SDO/AIA:][]{lemen2011}.  The corresponding WCS coordinates from the SDO/AIA header are used here. 

\begin{figure}   
\centering
\includegraphics[width=0.975\textwidth,clip=]{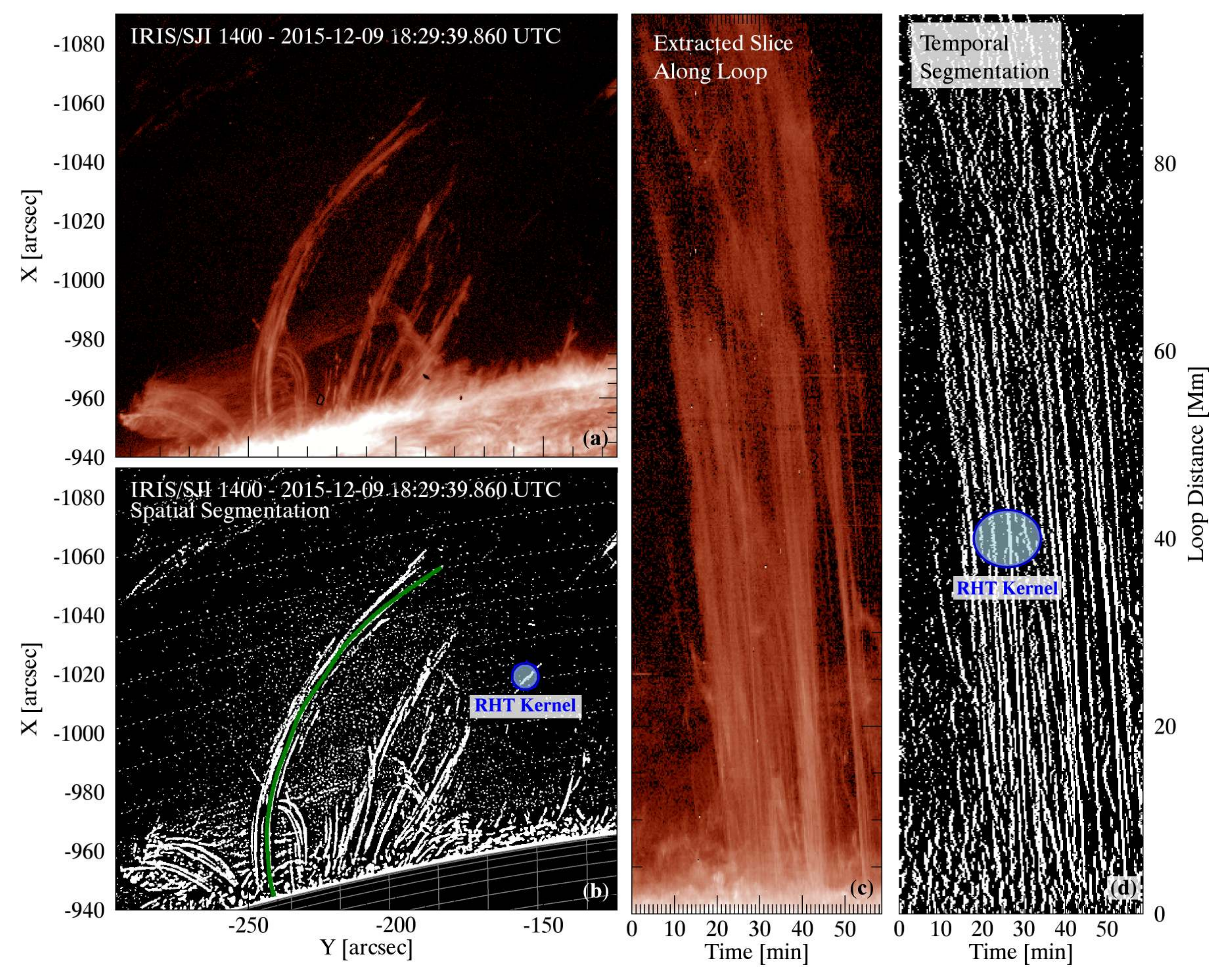}  

\caption{Spatiotemporal segementation of coronal rain data. \textit{(a)} Single IRIS slit-jaw 1400 \mbox{\AA} image during hour sequence with logarithm intensity scaling.  The horizontal and vertical axes display the `Y' and `X' helioprojective coordinates, which are referenced to Sun center with `Y' being parallel to the solar meridian. \textit{(b)} Spatiallly segmented version of image in panel (a) resulting from high-pass filtering the running mean (see text for details).  The green line denotes the manually traced loop used to demonstrate the temporal segmentation in panels (c) and (d).  The blue circle denotes the kernel window sized used for the spatial rolling hough transform.  \textit{(c)}  Extracted time-slice along the green line in (b). \textit{(d)} Temporally segmented version of (c) resulting from the bidirectional zero-phase difference filter.  The blue circle denotes the spatiotemporal kernel window used for the spatiotemporal rolling hough transform.}
\label{fig:iris_obs}	
\end{figure}

A snapshot extracted from the $1400$ $\mbox{\AA}$ IRIS/SJI time series is shown with logarithmic intensity scaling in Figure~\ref{fig:iris_obs}.  The region contains an active region prominence reaching heights approximately 20 Mm ($\approx 28''$; the physical scale is 714 km arcsec$^{-1}$) above the solar limb.  It erupted on 10 December 2015 at 3:50 UTC.  Multiple episodes of coronal rain were produced by the region within 24 hours prior to these IRIS observations.  The episode observed here is in progress at the beginning of the observation.  Numerous rain-producing coronal loops extend up to projected heights of 80 Mm above the region.  Other loop segments without footpoints in the observed field of view are evident at projected heights between 80 and 100 Mm.  Data counts in the raining material primarily lie below 30 (most below 15) for an individual image while the background noise standard deviation is $\sigma_{noise} \approx 1.1$. 

\subsubsection{Segmentation and RHT parameters}

The procedure outlined above is applied to only the off-limb portion of the observed field-of-view.  A temporal mean of the time series is first calculated using a centered kernel of width $w_{r} = 10$ ($\approx 3.2$ minutes) so that the coronal rain blobs evolve sufficiently to outline the projected trajectories.  As the blobs are by their nature elongated along their paths, this step significantly enhances the signal to noise of the raining features.  Next, each frame is highpass filtered by subtracting a boxcar averaged smoothed version of itself using a boxcar width of 10 pixels ($\approx 1.66''$).  Finally, a spatially segmented version of the data cube is obtained by binarzing according to a threshold of 0.17 (equivalent to $\sigma_{noise} / (2\sqrt{w_{r}})$). An example showing the spatial segmentation is given in Figure~\ref{fig:iris_obs} (b). 

For temporal filtering and segmentation, a zero-phase-lag bidirectional difference filter is first applied to the original time series using a difference width of 2 time steps as follows:
\begin{align}
I_{bdf}(x,y)|_{k} &= (I(x,y)|_{k} - I(x,y)|_{k-2} ) - (I(x,y)|_{k} - I(x,y)|_{k+2}) \\
 	  &= 2I(x,y)|_{k} - I(x,y)|_{k-2} - I(x,y)|_{k+2}
\end{align}
where $I(x,y)|_{k}$ corresponds to the $k$th frame of the time series.  Some random noise is then filtered out by masking the data based on its 3D boxcar average using a cube width of 3 pixels.  Any pixels within the bidirection difference filtered data whose 3D boxcar average is less than 0.018 (equivalent to 0.5 count per $3 \times 3 \times 3$ kernel) are set to zero.  Finally, the result is segmented by binarizing according to a threshold of 0.  To demonstrate the result of this temporal segmentation, a time slice along a raining coronal rain is manually extracted (see green curve in Figure~\ref{fig:iris_obs} (b)) and shown in Figure~\ref{fig:iris_obs} panels (c) and (d).  This mirrors the process that the hierarchical RHT approach automatically performs when analyzing the data.  As illustrated, the coronal rain flows are well segmented in time and thus provide adequate features for the application of the temporal component of the RHT analysis. 

Based upon the estimated widths of the raining coronal loops ($<\approx 1''$), the RHT kernel width used to extract the spatial orientation of the apparent motion trajectories and the temporal orientation (\textit{i.e.} projected speeds) of the time-slice curves is $D_{W} = 51$ pixels.  In the spatial domain, this corresponds to a circle with a diameter of $(D_{W}-1) = 8.3''$.  In the spatiotemporal cross-section resulting from a slice along a feature, the kernel diameter is $8.3''$ along the spatial direction and $15.83$ minutes along the time axis.  The RHT kernel size is illustrated for the two cases in Figure~\ref{fig:iris_obs} panels (b) and (d).  


\begin{figure}   
\centering
\includegraphics[width=0.49\textwidth,clip=]{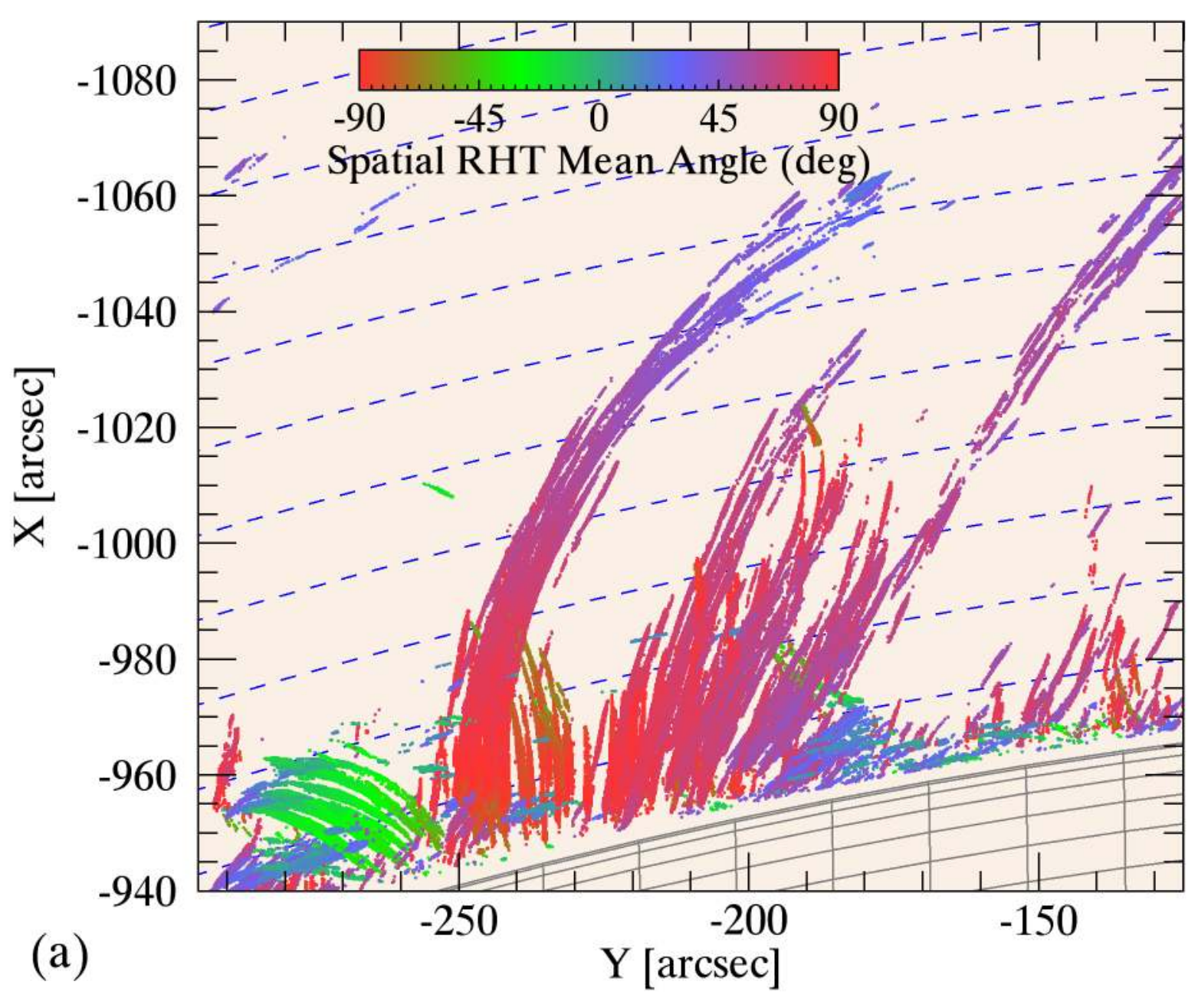} 
\includegraphics[width=0.49\textwidth,clip=]{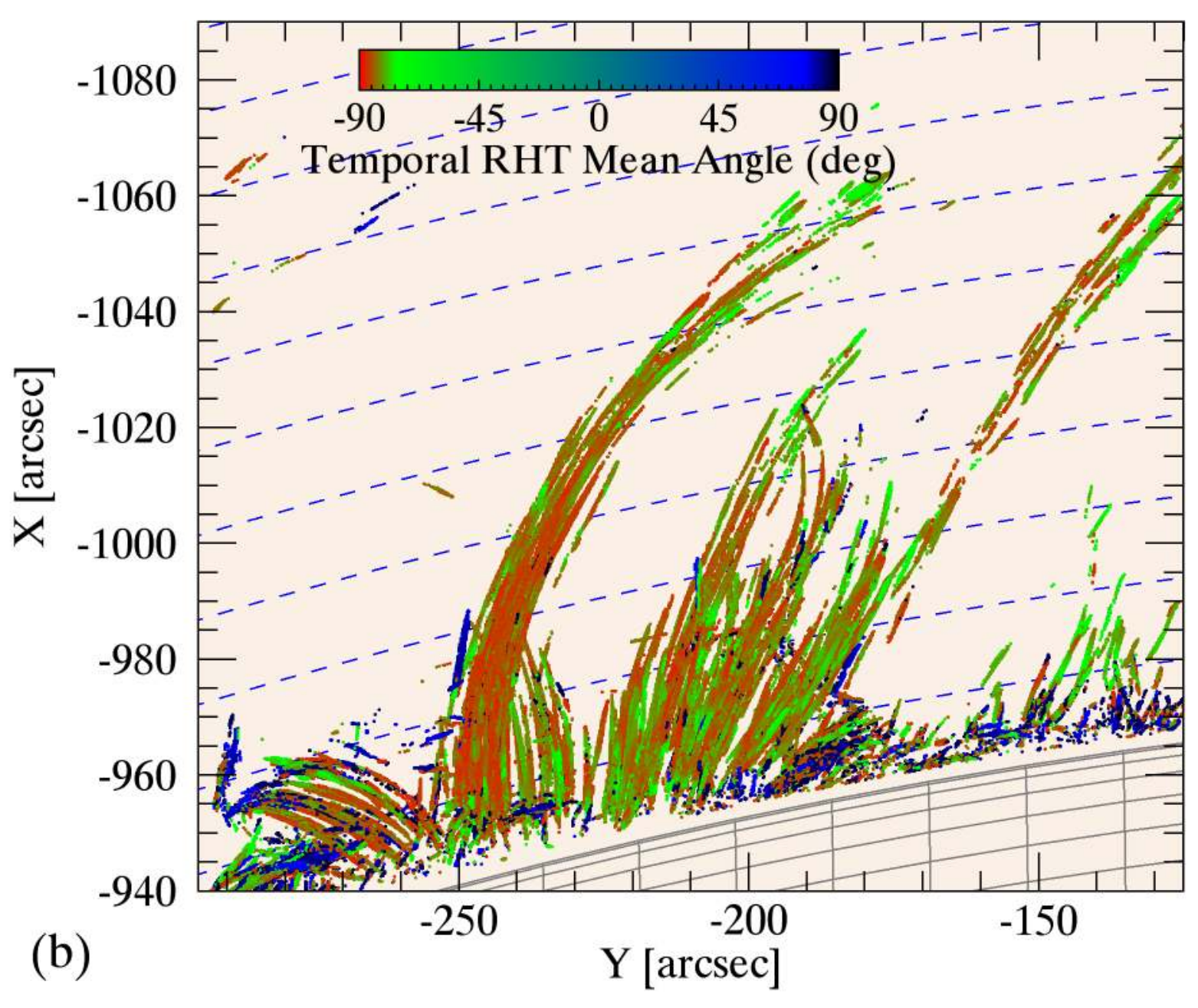} \\
\includegraphics[width=0.99\textwidth,clip=]{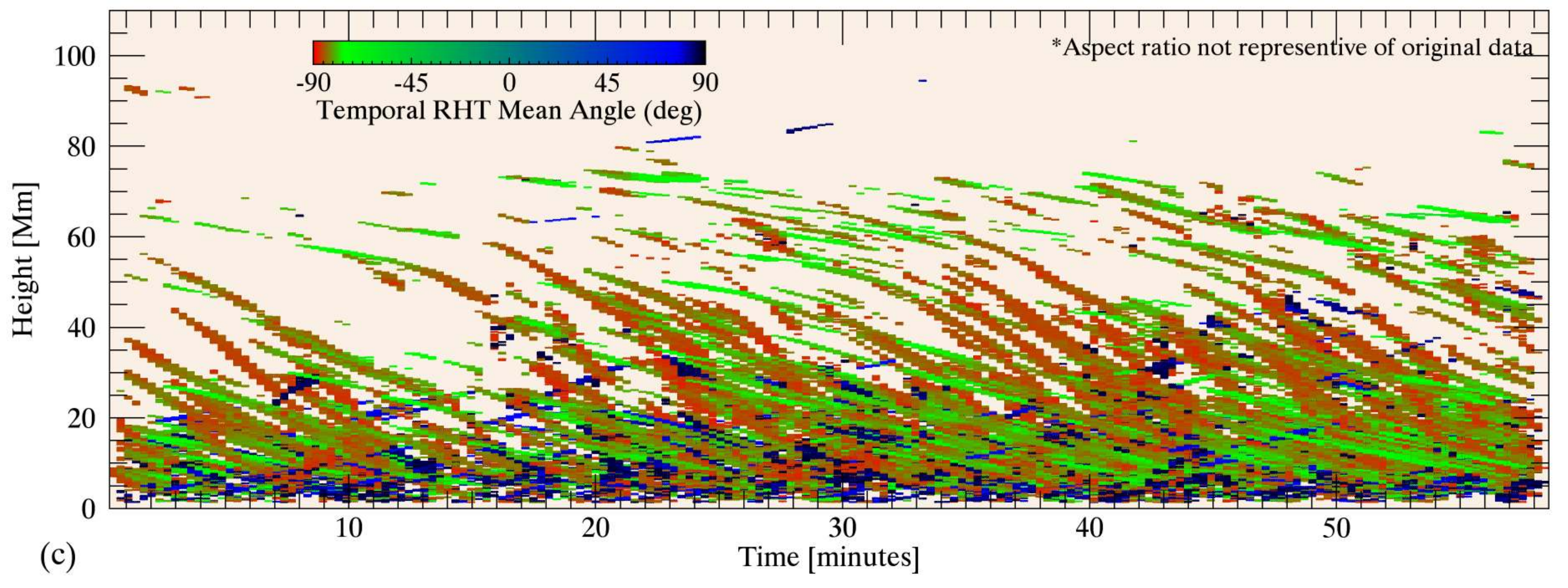} \\
\includegraphics[width=0.49\textwidth,clip=]{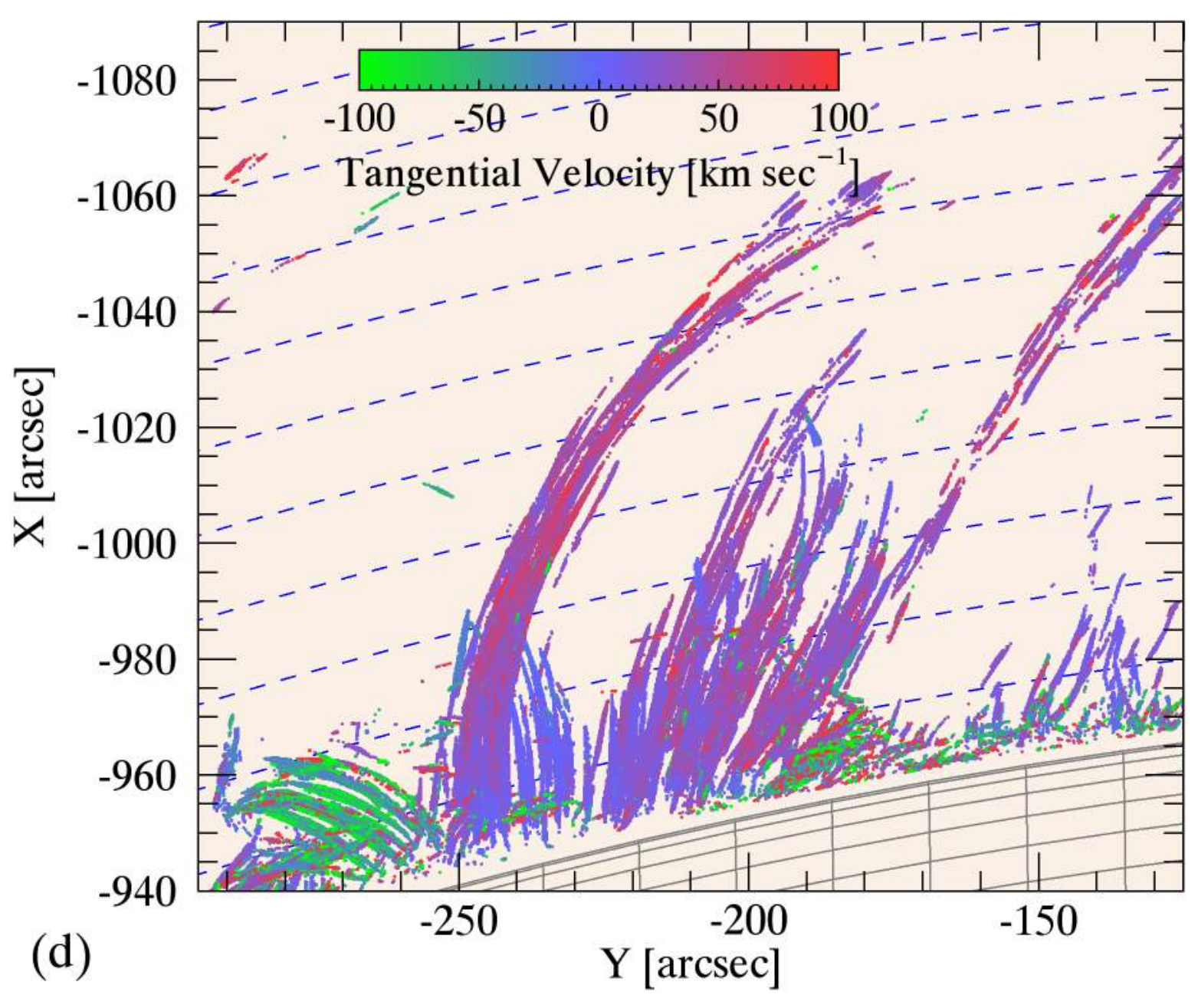} 
\includegraphics[width=0.49\textwidth,clip=]{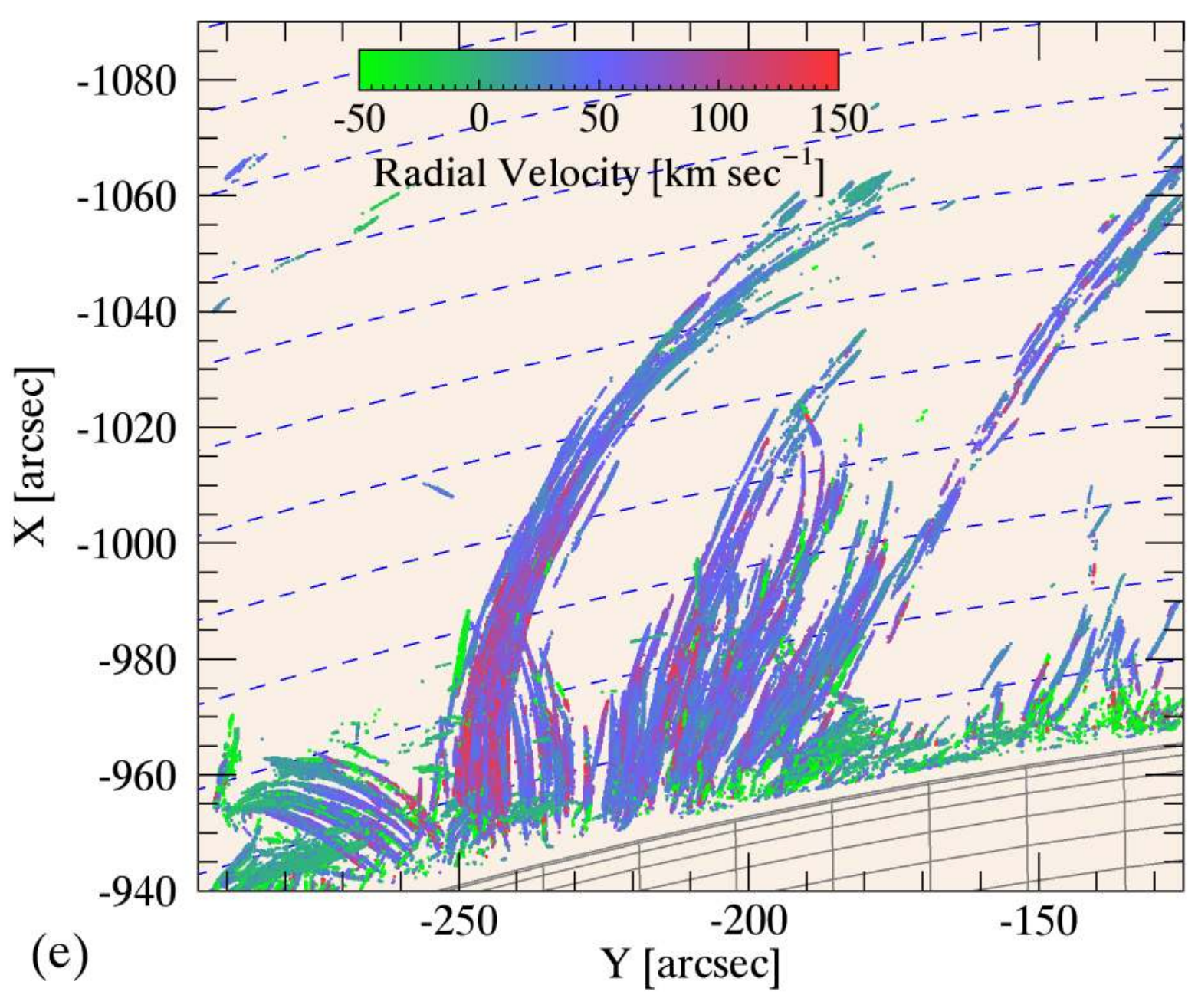} \\
\caption{Results of the application of the hierarchical RHT on coronal rain observed by IRIS.  (\textit{a}) Spatial maps of the mean spatial orientation angle for the raining coronal loops as projected in the plane of the sky.  The fiducial reference for the spatial RHT angle is aligned with the horizontal axis and increases in a counter-clockwise direction.  (\textit{b}) Spatial maps of the mean orientation angle in the spatio-temporal domain extracted along the projected spatial orientation of each feature, which is used to measure feature velocity. (\textit{c}) The spatio-temporal orientation angle derived from the RHT for all pixels and projected in the space-time plane (height vs time).  Note that the aspect ratio of this panel is not representative of the original data.  (\textit{d})  Spatial maps of the tangential component (wrt the solar limb) of the coronal rain velocity measured by the RHT.  Positive velocities are directed to the left in the panel. (\textit{e}) Spatial maps of the radial velocity component with positive values corresponding to flows directed towards the solar disk.}
\label{fig:iris_results}	
\end{figure}

\subsubsection{Results}

The results of the hierarchical RHT analysis are displayed in Figure~\ref{fig:iris_results}.  Only pixels for which $\bar R_{xy} \ge 0.9$, $\max[ H_{xy}(\theta) ] \ge 0.8$, $\bar R_{t} \ge 0.9$, and $\max[ H_{t}(\theta) ] \ge 0.8$ are included.  Approximately 10x more pixels are found if the $\bar R_{t}$ and $\max[ H_{t}(\theta) ]$ criteria are not considered, which is a result of the temporal averaging of the data used ($w_{r} = 10$) to find the projected angle of the apparent motion in the plane of the sky.  Since these pixels represent the 2D projection of a 3D trajectory, not all lie on the 3D spatiotemporal curve itself. The four criteria above together locate pixels on 3D trajectories with significant directionality.  In addition, any pixel with $\left | \bar\theta_{t}\right | \ge 88^{\circ}$ are removed as the resulting velocity error is very large, as limited by the temporal resolution of the data.  $2.9 \times 10^{5}$ off-limb pixels ($0.16\%$ of the data cube) remain after filtering.

The projected spatial orientation of the rain material, \textit{i.e.}, the spatial RHT mean angle $\bar\theta_{xy}$, is shown in Figure~\ref{fig:iris_results}(a) while the temporal RHT mean angle $\bar\theta_{t}$ is shown in Figure~\ref{fig:iris_results}(b).  All $2.9\times 10^{5}$ points have unique $(x,y,t)$ coordinates within the data cube but are here shown in projection on the plane of the sky; thus, many points overlap with other points.  Compare these plots with that of Figure~\ref{fig:iris_obs}(a) and note that many more loops are apparent in this projection as it represents the full time series (not a snapshot).  The reference direction for the spatial RHT angle is aligned with the horizontal axis and increases in a counter-clockwise direction.  As is visibly apparent in Figure~\ref{fig:iris_results}(a), the projected loop directions are well recovered by the spatial RHT, as expected based on the results in the previous sections. 

The reference direction for the temporal RHT angle $\bar \theta_{t}$ is aligned with the temporal axis and has positive values for apparent motion directed upwards in Figure~\ref{fig:iris_results}(b), \textit{i.e.}, towards more negative X helioprojective angles.  As is shown in panel (b), the temporal RHT results are dominated by downward motion (negative values of $\bar \theta_{t}$) as expected for the coronal rain event.  Furthermore, most of values are less than $-45^{\circ}$, meaning the projected velocities are primarily greater than $\approx 6.2$ km sec$^{-1}$, \textit{i.e.}, $(0.166'' \times 714$ km arcsec$^{-1})/ 19$ sec.  The color table in panel (b) emphasizes angle variations between $-45^{\circ}$ and $-90^{\circ}$ to bring attention to the height dependent behavior of $\bar \theta_{t}$.  In particular for the curved structure with endpoints near $<Y,X> = <-250,-950>$ and $<Y,X> = <-180,-1050>$, its clear that $\bar \theta_{t}$ generally decreases (velocity increases) for points closer to the limb, which is consistent with the known downward acceleration of coronal rain. 

To further illustrate that the temporal RHT is able to extract differences in velocity along the apparent motion trajectories, Figure~\ref{fig:iris_results}(c) shows a space-time projection for the $\bar \theta_{t}$ values for all filtered results.  As before, in projection many points overlap with other points; however, many of the space-time curves of the raining material are identifiable.  Although the aspect ratio of this panel is stretched by a factor of $\sim12$ compared to the observational pixel scales to boost clarity, the apparent angles of the projected features do systematically follow, and are consistent with, the reported value of $\bar \theta_{t}$ (taking into account the aspect ratio). 

The velocity along each curvilinear feature $v_{\parallel}$ is derived from the temporal RHT angle and the data sampling rates as
\begin{equation}
v_{\parallel} = \left ( \frac{\delta x}{\delta t} \right )  \tan \bar \theta_{t} 
\end{equation}
where $\delta x = 119$ km (the IRIS spatial sampling on this date) and $\delta t = 19$ sec.  The horizontal and vertical components of the apparent velocity are found from the projected loop orientation as 
\begin{align}
v_{horz} &= v_{\parallel} \cos \bar \theta_{xy} \\
v_{vert} &= v_{\parallel} \sin \bar \theta_{xy},
\end{align}
from which the tangential ($v_{tan}$) and radial ($v_{rad}$) velocity components, with respect to the Sun, are derived.  Maps of the projected tangential and radial velocities are shown in Figure~\ref{fig:iris_results} panels (d) and (e) showing, once again, behavior consistent with the downward acceleration of coronal rain. 

\begin{figure}  
\centering
\includegraphics[width=0.975\textwidth,clip=]{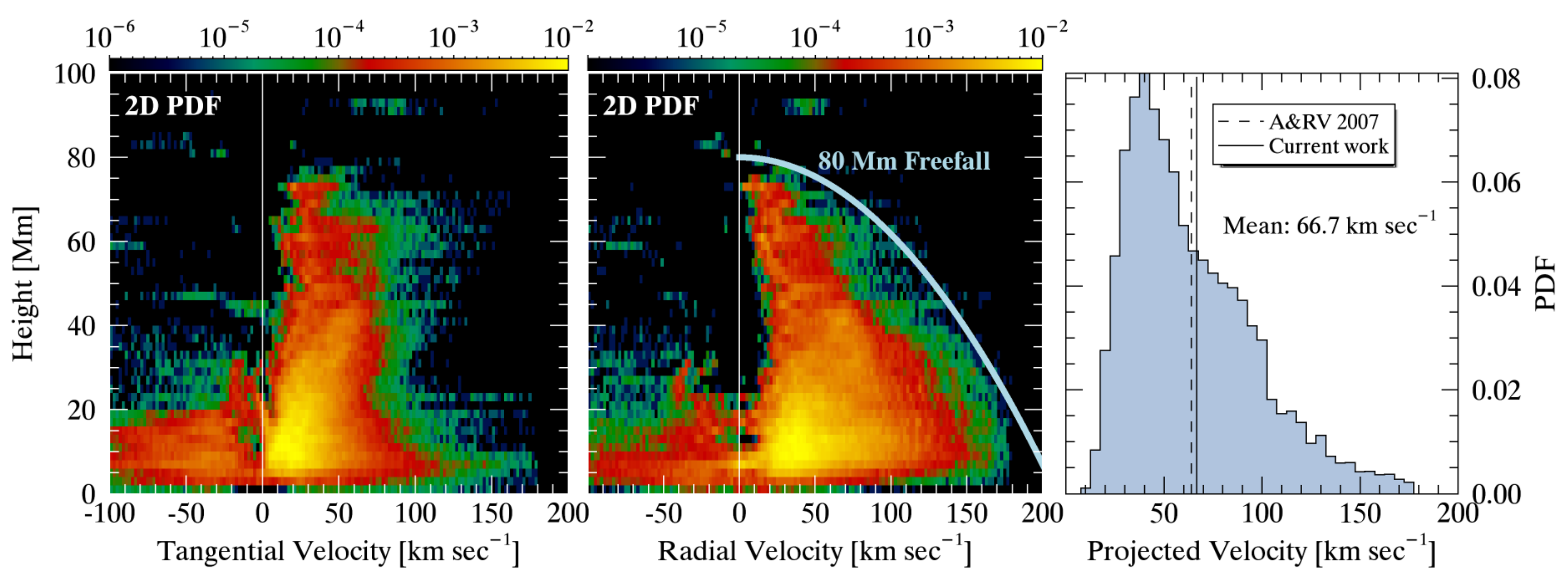} 
\caption{Measured probability distribution functions (PDFs) for coronal rain flows above NOAA AR 12468.  (\textit{left})  Two-dimensional PDF of the material height and the tangential (wrt to the solar limb) component of velocity projected in the plane of the sky.  Positive values of velocity are directed towards the south pole.  (\textit{middle}) Two-dimensional PDF of the material height and the radial component of velocity projected in the plane of the sky.  Positive values are directed towards the solar disk.  The light blue curve represents the free fall velocity curve for material raining from a height of 80 Mm with zero initial velocity. (\textit{right}) One-dimensional PDF of the total projected velocity ($\sqrt{(v_{tan}^{2} + v_{rad}^{2})}$.  The mean PDF value () is given by the solid vertical line and compared with that of \cite{antolin2012sharp}, i.e. A\&RV 2012.}
\label{fig:rain_vel_histograms}	
\end{figure}

For a first-cut analysis of the derived velocities, two-dimensional probability distribution functions (PDFs, or histograms) are calculated for the tangential and radial velocities relative to material height (Figure~\ref{fig:rain_vel_histograms}), as well as a 1D histogram of the total projected velocity ($\sqrt{(v_{tan}^{2} + v_{rad}^{2}}$).  The shape of the 2D PDF for the radial velocity forms an envelope of downward velocities consistent with the freefall velocity limit.  The overplotted white curve shows the height-dependent velocity of material undergoing freefall under solar gravity initially at a rest at a height of 80 Mm.  Nearly all of the 2D PDF power lies under this curve, which is consistent with and improves upons previous studies of coronal rain \citep[compare to Figure 6 of][]{antolin2012sharp}.  The 1D PDF of projected velocity also gives a mean value ($66.7$ km sec$^{-1}$) consistent with \citeauthor{antolin2012sharp} ($\approx 70$ km sec$^{-1}$) (labeled A\&RV in Figure~\ref{fig:rain_vel_histograms}); although, the shape of the PDF is somewhat different, likely due to variations in the regions studied and/or greater sensitivity here to the detection of slower moving events. 

\begin{figure}   
\centering
\includegraphics[width=0.5\textwidth,clip=]{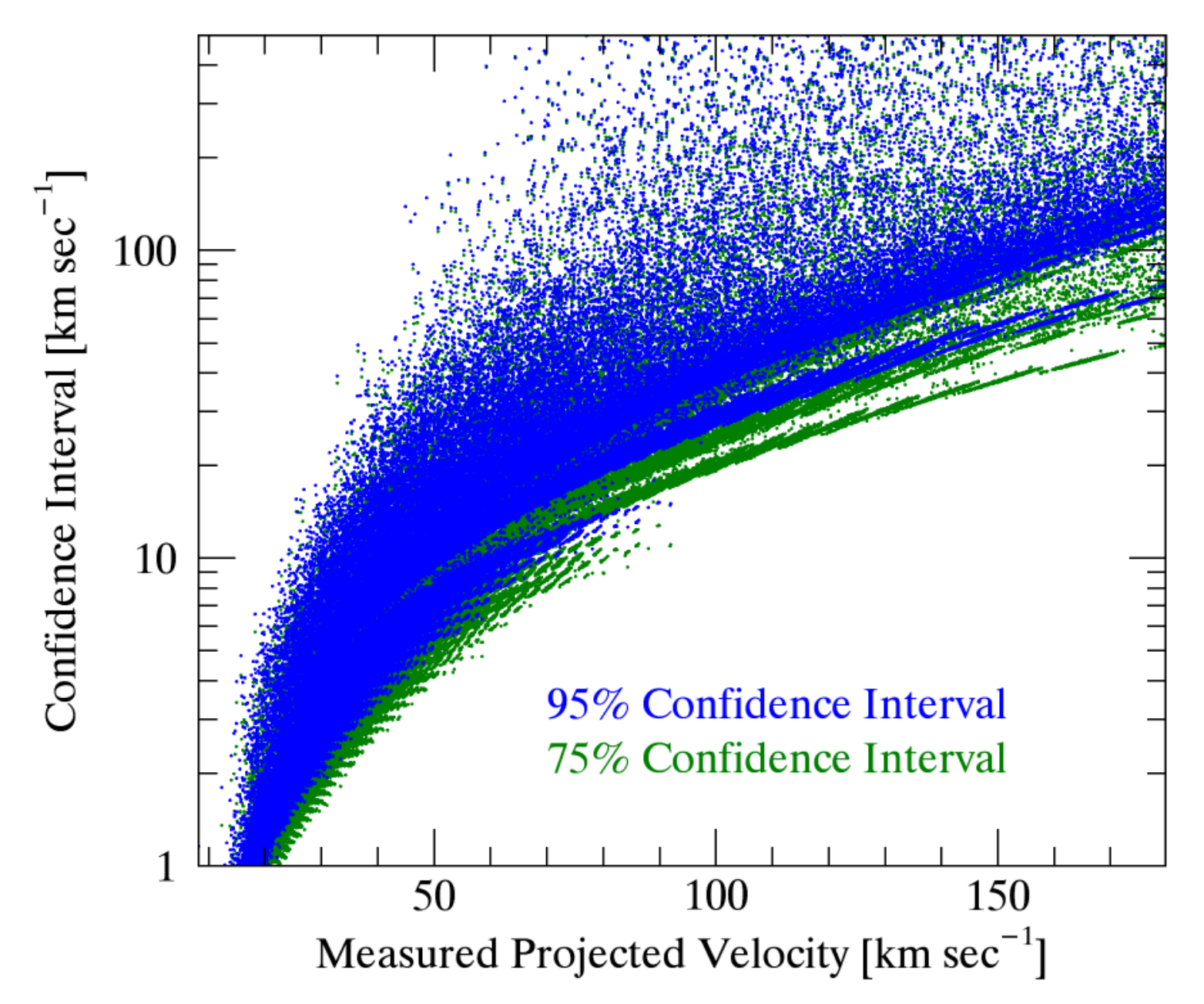} 
\caption{Confidence intervals for the projected velocity scale with its amplitude due to the limit imposed by the temporal resolution of the original data.} 
\label{fig:rain_vel_errors}	
\end{figure}

Unlike manual time-slice analysis where error estimation is difficult, Equation~\ref{eqn:conf_int} gives the confidence interval for the derived RHT angles and therefore error estimates for the derived projected velocities.  The angular confidence interval results in different lower and upper bounds for the velocity errors; however, to summarize the errors, we calculate the average confidence interval according to
\begin{equation}
\delta v_{\parallel} \approx \left ( \frac{\delta x}{\delta t} \right ) \frac{
	\left |   \tan \left ( \bar \theta_{t} + \epsilon_{\theta}|_{t} \right ) - 
		     \tan \left ( \bar \theta_{t} - \epsilon_{\theta}|_{t} \right ) \right | }{2}.
\end{equation}
In Figure~\ref{fig:rain_vel_errors}, the average 95\% confidence interval for the total projected coronal rain velocities is plotted as a function of total projected velocity.  As expected due to the limited temporal resolution of the data, the errors scale with the total projected velocity.  If a reasonable maximum cutoff for measurable projected velocities is defined as $(\delta x / \delta t) \tan (\pi/2. - \delta \theta)$, where $\delta \theta$ is the angular sampling of the RHT given in Equation~\ref{eqn:theta_sampling}, then here the maximum cutoff is $\approx 311$ km sec$^{-1}$.  Due to the large errors for high velocities, the results shown here are limited to those below 178 km sec$^{-1}$ by requiring that  
$\left | \bar\theta_{t}\right | \le 88^{\circ}$.  Considering the congruence of the Figure~\ref{fig:iris_results} with the freefall envelope, the 95\% confidence interval may be an overly conservative estimate of the errors.  For this reason, the 75\% confidence intervals are also shown in Figure~\ref{fig:rain_vel_errors}.

\section{Discussion}

The RHT is a useful technique for quantifying curvilinear directionality within solar imaging data sets.  This work has extended the concept of the RHT by first formulating an approach to RHT error analysis and then developing procedures to use the RHT in dense quasi-continuous regimes (chromospheric fibrils) and in multi-dimensional data sets (coronal rain).  The coronal rain application relies upon a hierarchical RHT approach that automates time-slice analysis for apparent motion quantification.  In each use case, the extended RHT performs well and the error analysis is able to properly filter the results.  For the coronal rain observations, the derived velocities show consistency with previous results, which provides confidence in the technique.  

Some key advantages of this RHT implementation have already been discussed including its ability to quantify directionality for intra-ridge pixels,  its low signal-to-noise detection capability, and the ability to remove pixels without sufficient linear coherence based on the mean resultant length (\textit{i.e.}, circular dispersion).  Another advantage is the numerical simplicity of the RHT and therefore its speed.  RHT analysis of the TRACE, IBIS, and IRIS data sets above, respectively, required 6 seconds, 14 minutes, and 1.5 hours of computation time; however, these are considered upper bounds on time requirements as little effort was made in code optimization.\footnote{Primarily single-core operations within the Interactive Data Language (IDL$^{\tiny\textregistered}$) were used on a 3.3 GHz processor accessing DDR4 memory. IDL$^{\textregistered}$ is a product of Exelis Visual Information Solutions, Inc., a subsidiary of Harris Corporation (Exelis VIS).}  Run time scales with the number of pixels being analyzed and the kernel size. 

This RHT implementation has a number of limitations as well.  Currently, the hierarchical RHT time-slice procedures have been tuned on stationary curvilinear features for which apparent motion is directed along their axes.  It is not expected to perform well for cases where the feature exhibits additional motion traverse to its axis, as might be the case for erupting filaments, for example.  While the RHT can be used to assess directionality within such structures at each snapshot in time, automated time-slice analysis is less straightforward.

Feature \textit{tracing} and identifying individual curvilinear features in time or space by associating their corresponding pixels is also not an operation performed by the RHT, as has already been discussed.  The goal here has only been to use the RHT to evaluate feature orientation on a pixel-by-pixel basis, and not to, \textit{e.g.}, count the number of loops or extract feature lifetimes.  It is, however, expected that the RHT orientation results might be used as a starting point in iterative methods that make use of direction information to trace individual loops (as discussed in the introduction), which could be extended to spatio-temporal traces of dynamic features. 

A number of ways to extend the techniques developed here might be considered.  To keep the analysis simple and rapid, multimodal distributions, which result from intersecting features, have been ignored here.  Instead, pixels at intersections are largely filtered out based on the mean resultant length of their transformed function $H_{xy}(\theta)$. A multimodal RHT analysis may provide useful information on the continuity of features as they intersect with other features.  In addition, all calculations have been performed with a circular kernel of a singular size in order to simplify the statistical analysis.  However, this choice may not be numerically optimal in all cases, \textit{e.g.}, when the temporal cadence of the observations is very high or when there are multiple feature scales in the images.  An elliptical RHT kernel shape and/or an adaptive kernel size may be useful in such cases if the statical analysis techniques are developed further.  Similarly, its important to note that the RHT does not have to be limited to the analysis of orientations approximated as linear directions.  The approach can be generalized, as the Hough transformed often is, to other line shapes--\textit{e.g.} higher order polynomials used to measure coronal rain accelerations--but comes at the price of statistical and numerical complexity. 


\section{Summary}

In summary, the RHT algorithms adapted and developed here are robust techniques that successfully automate the direction characterization of curvilinear features in multidimensional data sets.  The ability to derive the orientation of coronal loops and chromospheric fine structure has been demonstrated and is expected to aid in the modeling of magnetic fields in the solar atmosphere.  In addition, a novel new approach to automated apparent motion analysis of off-limb coronal rain has been developed that greatly simplifies the statistical characterization of such flows. 


\acknowledgments

The National Solar Observatory (NSO) is operated by the Association of Universities for Research in Astronomy, Inc. (AURA), under cooperative agreement with the National Science Foundation.  IRIS is a NASA small explorer mission developed and operated by LMSAL with mission operations executed at NASA Ames Research center and major contributions to downlink communications funded by ESA and the Norwegian Space Centre.  The author is grateful to Kevin Reardon for providing the IBIS data set as well as for a careful reading of the manuscript.


\bibliographystyle{spr-mp-sola-limited}
\bibliography{schad_manuscript}


\end{article}

\end{document}